\documentclass[fleqn,usenatbib]{mnras}

\usepackage{newtxtext,newtxmath}
 
\usepackage[T1]{fontenc}

\DeclareRobustCommand{\VAN}[3]{#2}
\let\VANthebibliography\thebibliography
\def\thebibliography{\DeclareRobustCommand{\VAN}[3]{##3}\VANthebibliography}

\usepackage{graphicx}	% Including figure files
\usepackage{amsmath}	% Advanced maths commands
\usepackage{amssymb}	% Extra maths symbols
\usepackage{tabularx}
\usepackage{multirow}
\usepackage{makecell}
\usepackage{mathtools}

\title[M3: Calibration and Imaging]{Mining Mini-Halos with MeerKAT I. Calibration and Imaging}

\author[K. S. Trehaeven et al.]{K. S. Trehaeven$^{1}$\thanks{E-mail: ktrehaeven@gmail.com},
	V. Parekh$^{1,2,3}$,
	N. Oozeer$^{2,4,1}$,
	B. Hugo$^{2,1}$,
	O. Smirnov$^{1,2}$,
	G. Bernardi$^{1,2,5}$,
	K. Knowles$^{1,2}$,
	\newauthor{C. Tasse$^{6,1}$,
		K. M. B. Asad$^{1,2,7,8}$,
		and S. Giacintucci$^{9}$}
	\\
	$^{1}$Department of Physics and Electronics, Rhodes University, PO Box 94, Grahamstown, 6140, South Africa\\
	$^{2}$South African Radio Astronomy Observatory, 2 Fir Street, Black River Park, Observatory, Cape Town, 7405, South Africa\\
	$^{3}$National Radio Astronomy Observatory (NRAO), 1003 Lopezville Rd, Socorro, NM 87801, USA\\
	$^{4}$African Institute for Mathematical Sciences, 6 Melrose Road, Muizenberg, 7945, South Africa\\
	$^{5}$INAF – Istituto di Radioastronomia, via P. Gobetti 101, 40129 Bologna, Italy \\
	$^{6}$GEPI \& USN, Observatoire de Paris, CNRS, Université Paris Diderot, 5 place Jules Janssen, 92190 Meudon, France\\
	$^{7}$ARGI, Department of Physical Sciences, Independent University, Bangladesh, Bashundhara RA, Dhaka, 1229, Bangladesh\\
	$^{8}$Department of Physics and Astronomy, University of the Western Cape, Bellville, Cape Town 7535, South Africa\\
	$^{9}$Naval Research Laboratory, 4555 Overlook Avenue SW, Code 7213, Washington, DC 20375, USA}

\date{Accepted 2023 January 31. Received 2023 January 23; in original form 2022 October 3}

\pubyear{2023}

\begin{document}
	\label{firstpage}
	\pagerange{\pageref{firstpage}--\pageref{lastpage}}
	\maketitle
	
	\begin{abstract}
		Radio mini-halos are clouds of diffuse, low surface brightness synchrotron emission that surround the Brightest Cluster Galaxy (BCG) in massive cool-core galaxy clusters. In this paper, we use third generation calibration (3GC), also called direction-dependent (DD) calibration, and point source subtraction on MeerKAT extragalactic continuum data. We calibrate and image archival MeerKAT L-band observations of a sample of five galaxy clusters (ACO 1413, ACO 1795, ACO 3444, MACS J1115.8+0129, MACS J2140.2-2339). We use the CARACal pipeline for direction-independent (DI) calibration, DDFacet and killMS for 3GC, followed by visibility-plane point source subtraction to image the underlying mini-halo without bias from any embedded sources. Our 3GC process shows a drastic improvement in artefact removal, to the extent that the local noise around severely affected sources was halved and ultimately resulted in a 7\% improvement in global image noise. Thereafter, using these spectrally deconvolved Stokes I continuum images, we directly measure for four mini-halos the flux density, radio power, size and in-band integrated spectra. Further to that, we show the in-band spectral index maps of the mini-halo (with point sources). We present a new mini-halo detection hosted by MACS J2140.2-2339, having flux density $S_{\rm 1.28\,GHz} = 2.61 \pm 0.31$ mJy, average diameter 296 kpc and $\alpha^{\rm 1.5\,GHz}_{\rm 1\,GHz} = 1.21 \pm 0.36$. We also found a $\sim$100 kpc southern extension to the ACO 3444 mini-halo which was not detected in previous VLA L-band observations. Our description of MeerKAT wide-field, wide-band data reduction will be instructive for conducting further mini-halo science.
	\end{abstract}
	
	
	%%%%%%%%%%%%%%%%%%%%%%%%%%%%%%%%%%%%%%%%%%%%%%%%%%
	
	%%%%%%%%%%%%%%%%% BODY OF PAPER %%%%%%%%%%%%%%%%%%
	
	\section{Introduction}
	
	Galaxy clusters are the largest virialised structures in the Universe. They form at the intersection of the Cosmic Web filaments and often grow by capturing smaller systems (for example, galaxies or groups) in their gravitational potential well, in so-called merger events. Clusters host most of their baryonic mass in the hot, tenuous, intra-cluster medium (ICM, all abbreviations used in this work are listed in Table \ref{tab:Table A.1} of Appendix \ref{appendix: abbreviations}), in which merger events deposit most of their energy. The dynamical state of the ICM can be probed in several ways, for example through thermal bremsstrahlung radiation observed in X-rays, as well as via the presence of diffuse radio emission - which is observed in an ever-increasing number of clusters \citep{2012A&ARv..20...54F}. This faint, diffuse emission is caused by a pool of non-thermal ultra-relativistic electrons emitting synchrotron radiation in the presence of intra-cluster magnetic fields. A combined analysis of the X-ray and radio cluster environment can inform on the relationship between its thermal and non-thermal components.
	
	Extended radio emission found within galaxy clusters generally has a low surface brightness and is classified into three broad phenomena, namely radio relics, giant radio halos and radio mini-halos; see \citet{2019SSRv..215...16V} for a comprehensive observational review of diffuse radio emission from galaxy clusters. Specifically, radio mini-halos are generally a few 100 kpc in size, confined to the cool-core of relaxed (non-merging) massive (> $5\times10^{14} M_{\odot}$) clusters, typically have irregular morphologies, surround the Brightest Cluster Galaxy (BCG), and are unpolarized \citep{2014ApJ...781....9G,2014ApJ...795...73G}. They are difficult to image and just over 40 mini-halos (including candidates) are known to date \citep{2022A&A...657A..56K,2022arXiv220201235B} with many more predicted for future discoveries \citep{2018A&A...617A..11G}. The origin of the synchrotron-emitting electron population is still under debate, however, it is thought to be closely linked to the cluster thermal component and/or the BCG. The particle (re)-acceleration mechanism responsible for the production of mini-halos is argued to either be turbulent (re)-acceleration via cool-core gas sloshing \citep{2012AdAst2012E...6G,2013ApJ...762...78Z,2020MNRAS.499.2934R,2022MNRAS.512.4210R} and/or hadronic collision models \citep{2015ApJ...801..146Z,2017MNRAS.467.1449J,2017MNRAS.467.1478J}.
	%Radio relics are filamentary-like sources located at the periphery of dynamically disturbed clusters and are thought to trace the propagating shock waves of merger events \citep[e.g.][]{2005ApJ...627..733M,2013MNRAS.435.1061P}. Giant radio halos are centrally located Mpc-sized sources thought to arise from the turbulent re-acceleration in the ICM via merger events, and so roughly follow the X-ray emission of the host cluster \citep[e.g.][]{2001MNRAS.320..365B,2007MNRAS.378..245B}.
	
	%Most of the detected diffuse radio emission from galaxy clusters is from low- to mid-range redshifts (< 0.5), with the highest redshift halo being at z = 0.77 \citep{2020A&A...642A..70O}. Pushing the detections to higher redshifts may allow astronomers to study the time-evolution of such sources, but the difficulties in doing so include the low surface brightness and small angular scales that such sources reduce to at high redshifts. To overcome these challenges, the new generation of radio telescopes are designed to maximise sensitivity and resolution - for examples of deep MeerKAT continuum images see \citet{2022ApJ...925..165H,2022MNRAS.509.2150H}. 
	
	The MeerKAT telescope \citep{2016mks..confE...1J} located in the Karoo desert of South Africa, consists of 64 dishes, each 13.5 m in diameter resulting in a primary beam of just over one degree in diameter at L-band. Approximately three-quarters of its collecting area lies within a dense core region of diameter 1 km; the remaining dishes are spread to a maximum baseline of 7.6 km, giving a minimum angular scale of $\sim$5$\arcsec$ at L-band. This configuration allows MeerKAT to observe low surface brightness diffuse emission while simultaneously resolving compact sources. This is particularly important in the context of radio mini-halos, as the embedded Active Galactic Nucleus (AGN, which is often only a few arcseconds in size) needs to be disentangled from the diffuse synchrotron emission in order for a precise image and study to be made. Additionally, the low system temperature and high number of correlator frequency channels (in this case 4096) allows for the precise removal of radio frequency interference (RFI), increasing sensitivity to low surface brightness emission. 
	%parabolic dishes, all with a 13.5 m diameter and offset Gregorian optics. The majority of its collecting area (approximately 3/4th of the dishes) lies within a dense core region of diameter 1 km with shortest baseline of 29 m, and the remaining dishes spread to a maximum baseline of 7.6 km, giving a maximum and minimum angular scale of $\sim$0.5$^{\circ}$ and $\sim$5$\arcsec$ at L-band, respectively. The many short baselines allow MeerKAT to observe low surface brightness large-scale diffuse emission while the longer baselines resolve the compact emission. The low system temperature ($\sim$20 K) induces very little thermal noise to the receiver systems, allowing them to register very stable gain measurements. The large number of correlator frequency channels (in this case 4096 at the time of project proposal submission) allows for precise removal of radio frequency interference (RFI) from the observation data. All these attributes add to make MeerKAT an exceptionally sensitive telescope to both small- and large-scale continuum emission. This is particularly important in the context of radio mini-halos, because the embedded AGN (often only a few kpc in size) needs to be disentangled from the diffuse synchrotron emission in order to precisely image and study the mini-halo. 
	
	With radio telescopes becoming more powerful, calibration and imaging techniques have been developed to deal with the added complications of higher sensitivity data. The standard calibration techniques of cross-calibration (first generation calibration, or 1GC) and traditional self-calibration (second generation calibration, or 2GC) only correct for direction-independent effects (DIEs). However, the wide field-of-view (FoV) and large fractional bandwidth of modern radio telescopes significantly complicates the calibration problem by introducing considerable direction-dependent effects (DDEs), such as the atmospheric conditions in the ionosphere and the telescope's primary beam response \citep{2019MNRAS.485.4107I,2021MNRAS.502.2970A,2022AJ....163..135D}. The additional computational challenges limit the extent to which the resulting image artefacts can be addressed, thereby limiting overall image sensitivity. Novel direction-dependent (DD) calibration (hereafter referred to as third generation calibration, or 3GC) techniques that correct for the residual errors are being used to rectify this problem - for example, the faceted approach of killMS and DDFacet \citep{2014arXiv1410.8706T,2014A&A...566A.127T,2018A&A...611A..87T} or the source `peeling' of CubiCal \citep{2018MNRAS.478.2399K}. Implementing such techniques on modern telescopes such as the MeerKAT allows for images of higher dynamic range (DR - a common measure of the depth or quality of a continuum image) to be produced, thereby unveiling emission that was otherwise undetectable in previous observations.
	
	Many science cases require the subtraction of unwanted compact sources, such as the study of faint transient objects in the image plane \citep[for example][]{2022arXiv220710973G} and, of course, the study of extended diffuse sources in which compact emission is blended \citep[as in][]{2019A&A...622A..24S}. Source subtraction is particularly important for mini-halo studies, as a significant fraction of the mini-halo area is contaminated by the embedded AGN, which should be removed to avoid biasing any physical measurements. The effectiveness of such a technique is difficult to quantify, and interestingly, 3GC adds complexity to the issue which has not been explored in much detail. It is known that 3GC can worsen the effect of flux suppression/absorption; flux from faint unmodelled sources can be absorbed by the antenna gains and often transferred to the brighter modelled sources, thereby decreasing the observed flux density of the faintest sources and increasing that of the brighter ones \citep[see for example][]{10.1093/mnras/stu268,10.1093/mnras/stw118,10.1093/mnras/stz3037}. Thus, we note that examining the effect of 3GC on source subtraction could prove to be a useful exercise, however this is outside the scope of our current research.
	%We, however, simply provide a few image statistics to show that DD calibration is compatible with source subtraction and note that a rigorous statistical study is outside the scope of this work.
	
	In this paper, we present the calibration and imaging of mini-halo results using MeerKAT archival L-band observations for a sample of five galaxy clusters. We provide a guideline on how to process MeerKAT radio data to improve image quality via facet-based 3GC and unambiguously study the diffuse emission after point source subtraction. We describe the sample in Section \ref{section:sample}. Thereafter, in Section \ref{section:Data processing}, we describe our data reduction procedures, including 3GC, source subtraction and spectral imaging. The results of the MeerKAT and mini-halo data processing are presented in Section \ref{section:results}, with a discussion thereof provided in Section \ref{section:discussion}. Finally, a summary and conclusions are given in Section \ref{section:summary}.
	
	Throughout this paper, we adopt the physically motivated mini-halo definition proposed by \citet{2017ApJ...841...71G}, which states that a mini-halo is a diffuse radio source at the cluster center that (1) does not consist of, or have any morphological connection to, any emission directly associated with any embedded galaxies, (2) has a minimum radius of approximately 50 kpc and (3) a maximum radius of 0.2R$_{500}$\footnote{R$_{500}$ is the radius that encloses a mean overdensity of 500 with respect to the critical density at the cluster redshift}. We adopt a flat $\Lambda$CDM cosmology with H$_{0}$ = 70 km s$^{-1}$ Mpc$^{-1}$, $\Omega_{m}$ = 0.3 and $\Omega_{\Lambda}$ = 0.7. Finally, we describe the spectral nature of the radio emission as $S_{\nu} \propto \nu^{-\alpha}$, where $S_{\nu}$ is the measured flux density at frequency $\nu$ and $\alpha$ is the spectral index.
	
	\section{Sample selection}
	\label{section:sample}

	\begin{table*}
		\caption{Physical properties of the cluster sample. Columns: cluster name, cluster redshift, angular-to-physical scale calculated for our cosmology, cluster mass taken from \citet{2014A&A...571A..29P} unless stated otherwise, core X-ray morphology, core radio morphology and literature references.\\ 
			CC = cool-core, nCC = non-cool-core, MH = mini-halo, cMH = candidate mini-halo. \\
			\textsuperscript{*} Taken from \citet{2011A&A...534A.109P}. \\
			\textsuperscript{**} New detection from this work.\\
			Ref.: (1) \citet{2019A&A...622A..24S}; (2) \citet{2017ApJ...841...71G}, (3) \citet{2018A&A...618A.152K}; (4) \citet{2019ApJ...880...70G}; (5) \citet{2020A&A...640A.108G}.}
		\centering
		\begin{tabular}{ccccccc}
			\hline
			Name                & z     & Scale     & $M_{\rm 500\,, SZ}$                 & X-ray             & Radio & Ref \\
			&       & (kpc/$\arcsec$)  & $(10^{14}M_{\odot}$)        & morphology        & morphology                               \\
			\hline
			ACO  1413           & 0.143 & 2.51      & 5.98$_{-0.40}^{+0.38}$      & nCC               & MH    & 1, 2 \\
			ACO  1795           & 0.062 & 1.20      & 4.54$\pm0.21$      & CC                & cMH   & 2, 3 \\
			ACO  3444           & 0.254 & 3.96      & 7.6 $_{-0.6}^{+0.5}$        & CC                & MH    & 2, 4 \\
			MACS J1115.8+0129   & 0.350 & 4.94      & 6.4 $\pm0.7$        & CC                & MH    & 2, 5 \\
			MACS J2140.2$-$2339   & 0.313 & 4.59      & 4.7\textsuperscript{*}      & CC                & MH\textsuperscript{**}     & 5 \\
			\hline
		\end{tabular}
		\label{tab:Table 1}
	\end{table*}

	\begin{table*}
		\caption{Observation details. Columns: cluster name, RA, Dec, observation date, observation time on-target, primary calibrator, secondary calibrator.}
		\centering
		\begin{tabular}{ccccccc}
			\hline
			Name                & RA               & DEC              &Observation date    & On target time &Primary               &Secondary \\
			& (h:m:s, J2000)   & (d:m:s, J2000)   &                    & (minutes)      & calibrator           & calibrator        \\          
			\hline
			ACO 1413            & 11:55:19.40      & 23:24:26.0       &  10 August 2019       & 117            & J0408$-$6545        & J1120+1420 \\
			ACO 1795            & 13:48:55.00      & 26:36:01.0       &  16, 17 June 2019     & 240            & 3C286             & 3C286 \\
			ACO 3444            & 10:23:54.80      & -27:17:09.0      &  20, 26 July 2019     & 228          & J0408$-$6545        & J1051$-$2023\\
			MACS J1115.8+0129   & 11:15:54.90      & 1:29:56.0        &  2 August 2019        & 93            & J0408$-$6545        & J1058+0133\\
			MACS J2140.2$-$2339   & 21:40:15.20      & -23:39:40.0      &  30 June 2019         & 87            & J1939$-$6342        & J2152$-$2828\\
			\hline
		\end{tabular}
		\label{tab:Table 2}
	\end{table*}
 
	A sample of five relaxed galaxy clusters was selected from the literature and observed by MeerKAT during its first Open Time Call in 2019 to detect and characterise the central diffuse radio emission in each (proposal ID: SCI-20190418-KA-01). Table \ref{tab:Table 1} summarizes the physical properties of each target. The clusters were chosen based on the classification of the core radio emission detailed in the literature at the time of proposal submission, having either a confirmed or candidate mini-halo, as well as the declination of the host cluster. ACO 3444 was a newly confirmed mini-halo when \citet{2019ApJ...880...70G} studied archival VLA narrow L-band observations. However, because of the MeerKAT's greater sensitivity to diffuse emission due to its denser coverage of short baselines, more diffuse emission may potentially be detected. Additionally, this cluster lies at an optimal observing declination for MeerKAT. ACO 1413, ACO 1795 and MACS J1115.8+0129 were the most southerly candidate mini-halos, having previous radio observations of insufficient quality to derive physical properties \citep[see][respectively]{2009A&A...499..371G,2014ApJ...781....9G,2016sf2a.conf..367P}; hence, new observations were proposed utilising the MeerKAT's angular resolution and sensitivity to confirm their nature. However, the first and last of which have since been confirmed by \citet{2019A&A...622A..24S} and \citet{2020A&A...640A.108G}, respectively. The last cluster in the sample, MACS J2140.2-2339, had no previous radio observations and was classified by \citet{2009ApJS..182...12C} to be a cool-core, massive cluster. Since \citet{2017ApJ...841...71G} studied a mass-limited sample of galaxy clusters and concluded that most of the massive, cool-core clusters possessed a mini-halo, this last cluster with its ideal southerly declination was identified to potentially be a new MeerKAT-mini-halo detection. The sample possesses a large declination range; two are near $\pm$25$^{\circ}$ each - the negative being near the ideal observing range of the MeerKAT and the positive being near that of its northern limit - and one near the celestial equator. This large declination range lends to the possibility of future synergy between MeerKAT/SKA-Mid and LOFAR in mini-halo studies \citep[ACO 1413 had already been observed by LOFAR in][and Riseley et al. in prep aims to provide an indepth MeerKAT+LOFAR study of the diffuse radio emission in this very interesting cluster]{2019A&A...622A..24S}. The sample varies in redshift from 0.06 to 0.35 and has mass $\gtrsim$ $5\times10^{14} M_{\odot} $.
	
	\section{Data Processing}
	\label{section:Data processing}
	
	All data were downloaded from the SARAO archive\footnote{\url{https://archive.sarao.ac.za}} - all specialized software used in the data reduction is listed in Table \ref{tab:Table B.1} of Appendix \ref{appendix: software}. The observations were conducted with full 4096 channelization, 856 MHz total bandwidth centred at 1283 MHz (L-band), and an 8 second dump rate - except for MACS J1115.8+0129 which had a dump rate of 2 seconds. Table \ref{tab:Table 2} summarizes the observation details. Primary calibrators were visited once for 10 minutes at the start of each observing block and were used for bandpass calibration and fluxscaling. Secondary calibrators and the target sources were then visited alternately for 2 and 15 minutes, respectively, where the secondary calibrator was used for delay and complex-gain calibration.
	
	The \texttt{KATDAL} (KAT Data Access Library) package was used during download to convert from MeerKAT Visibility Format (MVF) to the standard CASA \citep{2007ASPC..376..127M} Measurement Set (MS) while applying flags generated by the telescope's control and monitoring system. A known issue with early MeerKAT observations is that of poor calibrator positions\footnote{\url{https://skaafrica.atlassian.net/wiki/spaces/ESDKB/pages/1481015302/Astrometry}}. The sources J0408$-$6545 and J1939$-$6342 needed to be corrected by 0.52$\arcsec$ and 2.00$\arcsec$, respectively, which was done using CASA's \texttt{fixvis} task. Subsequent flagging, 1GC and 2GC were automated using the CARACal pipeline \citep{2020ascl.soft06014J}. First, the pipeline copies all sources into separate `calibrators' and `target' MS files. The calibrator data was flagged for the known problematic ranges of the spectral window, autocorrelations and any shadowed antennas using CASA's \texttt{flagdata} task. Thereafter, Tricolour \citep{2022arXiv220609179H} - an automated flagging software specifically developed for MeerKAT data - was used to flag persistent RFI with a particularly strict default flagging strategy from the CARACal MeerKAT files\footnote{\url{https://github.com/caracal-pipeline/caracal/blob/master/caracal/data/meerkat_files/stalin.yaml}} to ensure no erroneous data were carried into calibration. In order to improve image DR and the detection of faint diffuse emission, we perform 3GC on all data. Below we briefly describe our calibration process.
	
	\subsection{Cross-calibration (1GC)} 
	
	The calibrator sources were used to derive corrections for various instrumental and propagation eﬀects that corrupt the astrophysical signals (such as antenna gain fluctuations and atmospheric conditions). These corrections are applied to the target to mitigate the observational errors, which is known as cross-calibration. Firstly, CARACal modelled the primary calibrators using the MeerKAT local sky models (lsm) of each. In the case of 3C286\footnote{3C286 does not have a MeerKAT lsm yet}, we used CASA's \texttt{setjy}, externally from CARACal, to set the model using the polarisation information from \citet{2017ApJS..230....7P} and the updated polarimetry properties from Table 7.2.7 from the NRAO website \footnote{\url{https://science.nrao.edu/facilities/vla/docs/manuals/obsguide/modes/pol}}. The pipeline then derived delay, complex-gain and bandpass amplitude- and phase-corrections using CASA's \texttt{gaincal} and \texttt{bandpass} tasks. The bandpass solutions from the primary calibrator were applied to the secondary calibrator on-the-fly while its delay and complex-gain corrections were calculated. The derivation was repeated after a mild automated flagging. The final secondary gain solutions were scaled to that of the primary using CASA's \texttt{fluxscale} task. The complex-bandpass solutions of the primary calibrator and the delay and flux-scaled complex-gain solutions of the secondary were applied to the target data which was then frequency-averaged over five channels to save on storage space and future processing time. Finally, this cross-calibrated target data was flagged similarly to the calibrator data but with a slightly more relaxed flagging strategy\footnote{\url{https://github.com/caracal-pipeline/caracal/blob/master/caracal/data/meerkat_files/gorbachev.yaml}}.
	%Approximately 65-75\% of the target data was subsequently flagged.
	
	\subsection{Direction-independent self-calibration (2GC)}\label{section:2GC}
	
	2GC is the process of using the target data itself to improve the instrumental complex-gain corrections derived earlier and hence improve image DR. We used the CARACal defaults of WSClean \citep{2014MNRAS.444..606O} for imaging and CubiCal \citep{2018MNRAS.478.2399K} to calculate self-calibration solutions. Three rounds of phase-only self-calibration were performed. In each iteration, we imaged a 2.5$^{\circ}$ squared area spanning 6000 pixels (scale 1.5$\arcsec$/pixel) using multi-frequency synthesis \citep[MFS,][]{1990MNRAS.246..490C,1999ASPC..180..419S}, multi-scale cleaning \citep{2017MNRAS.471..301O}, \texttt{robust} 0 weighting \citep{1995AAS...18711202B}, a 2nd-order spectral polynomial fit, applying the built-in WSClean automasking routine (halving the masking threshold in each iteration, finishing on 3$\sigma$) and terminating the deconvolution after one million minor iterations. After each imaging round, we fed the model data into CubiCal to calculate phase-only self-calibrated corrected data. We solved for the gain solutions in 30 equal time chunks per calibration round over the entire bandwidth while flagging visibilities per each baseline and correlation using a median absolute residual (MAD) filter. Whilst the global image noise (root mean square of the residual image) steadily decreased, the reduction of severe artefacts from the 2nd to 3rd rounds of self-calibration was minimal; thus, 2GC was stopped at this point.
	
	\subsection{Direction-dependent self-calibration (3GC)}
	\label{section:3GC}
	
	3GC is the process of correcting the DD terms in the Radio Interferometer Measurement Equation (RIME) caused by DDEs \citep[see for example][for comprehensive discussions on generic DDEs]{2010A&A...524A..61N,2011A&A...527A.106S,2011A&A...527A.107S,2011A&A...527A.108S,2011A&A...531A.159S,2015MNRAS.449.2668S}. In the case of MeerKAT, the major effects are that of primary beam rotation and pointing errors. These manifest as radial artefacts around bright sources that lie towards the half-power point (flanks or sidelobes) of the telescope's primary beam\footnote{\url{https://skaafrica.atlassian.net/wiki/spaces/ESDKB/pages/1484128294/Dynamic+range+considerations}}. Instead of considering these sources on a case-by-case basis through the differential gain calibration or `source peeling' of CubiCal, we chose to calibrate over the entire field via the facet-based approach of killMS \citep[kMS, ][]{2014arXiv1410.8706T,2014A&A...566A.127T} and DDFacet \citep[DDF, ][]{2018A&A...611A..87T}. \citet{2022MNRAS.512.4210R} applied a similar approach to MeerKAT L-band observations of the MS 1455.0+2232 mini-halo. Additionally, we used the holographic MeerKAT L-band primary beam models generated via the \texttt{eidos} package to correct for the primary beam attenuation and rotation \citep{2021MNRAS.502.2970A,2022AJ....163..135D}. The kMS software performs 3GC on a tessellated DI image by exploiting Wirtinger complex differentiation \citep{2014A&A...566A.127T} to directly estimate the physical DD RIME terms. We used its non-linear Kalman Filter algorithm \citep[KAFCA, ][]{2014arXiv1410.8706T} to solve for stable and well-conditioned estimates within a 5 minute time and 10 channel frequency solution interval in as many directions as there were tessellation tiles. DDF is a facet-based imager that can take into account kMS solutions to perform wide-band, wide-field DD spectral deconvolution in the apparent and intrinsic (primary beam corrected) flux scale. Note that we describe all images in which primary beam corrections were performed as being intrinsic.
	
	The image tessellation mentioned above was determined using the \texttt{MakeModel.py} script\footnote{\url{https://github.com/saopicc/DDFacet/tree/master/SkyModel}} of the DDF software. Since 3GC is very sensitive to the signal-to-noise (SNR) of the data, we experimented with the number of tessellation tiles in each cluster case. We found that 4-6 tiles each centred on one of the brightest field sources responsible for radial artefacts gave a good improvement in image noise after 3GC. Once the tessellation pattern was defined, the 3GC procedure was analogous to that of 2GC (i.e. imaging a rigorous model then deriving the calibration solutions and applying them in a further imaging round).
	
	DDF offers a deconvolution algorithm called subspace deconvolution (SSD), which was shown in \citet{2018A&A...611A..87T} to work well in conjuction with kMS to produce 3GC intrinsic images. SSD jointly and independently deconvolves subsets/islands of pixels (defined using a mask) very efficiently, making it excellent for studying extended emission in large images. We used the SSD2 deconvolver on default settings (unless otherwise stated) throughout the 3GC imaging process - which simply adds to the original functionality to choose the order of the spectral polynomial fitted and a few other deconvolution parameters. 
	
	To produce the deepest DR images possible, imaging was split into an automasked initial deconvolution from which a rigorous mask was externally generated and applied in a direct continuation. Each initial deconvolution ran for two major cycles with a 5$\sigma$ automasking threshold, while each subsequent externally masked deconvolution ran for a single major cycle. After each deconvolution, an updated mask was generated with the tightest possible threshold (visually inspected to ensure only real sources were masked) and applied in the next step. The \texttt{breizorro} tool was used to generate all external masks, as it estimates a local noise rms so that a deeper mask can be achieved on smaller scales. The external masks had a threshold of 6-8$\sigma$ depending on the field, which could then be made even tighter after the 3GC solutions were applied. All images were produced using MFS with 10 sub-bands, \texttt{robust} 0 weighting, fitting a 4th-order spectral polynomial, superimposing a 20x20 square facet grid onto the tessellation pattern and re-evaluating the primary beam correction every 5 minutes in each facet to produce both apparent and intrinsic images throughout the process.
	
	The basic procedure is summarized as follows:
	\begin{enumerate}
		\item \textit{Tessellate the field;} 
		\item \textit{Initial deconvolution;}
		\item \textit{Deeper deconvolution;}
		\item \textit{Derive kMS DD gain solutions;}
		\item \textit{Apply kMS solutions in deconvolution;}
		\item \textit{Final deconvolution.}
	\end{enumerate}
	
	\subsection{Source subtraction}
	\label{section:src_sub}
	
	In most cases, the central BCG - which may have a complex morphology with jets/lobes - blends with the diffuse mini-halo emission. These sources need to be removed in order to obtain an uncontaminated and unbiased detection of the underlying mini-halo. Most modern mini-halo studies choose to perform visibility-plane subtraction \citep[for example][]{2021MNRAS.508.3995B,2022MNRAS.512.4210R}. This involves subtracting the modelled compact sources - which can be generated from a high-resolution (HR) image - directly from the visibilities and then re-imaging the residual data at the original resolution, thereby analysing the mini-halo without bias from the embedded sources. This approach introduces an additional uncertainty component into the measurement-taking process (Equation \ref{2}), which scales with the size and strength of the subtracted sources. We performed visibility-plane subtraction on the final 2GC and 3GC images produced in Section \ref{section:3GC}.
	
	All imaging, deconvolution and masking parameters outlined in Section \ref{section:3GC} were followed for the subtraction, except where otherwise stated. The kMS solutions derived in Step 4 of Section \ref{section:3GC} were applied throughout the subtraction procedure. HR deconvolutions were performed at \texttt{robust} -2 while applying a uv-cut to model only the compact emission present in the field, giving images at a $\sim$4$\arcsec$ resolution. The uv-cut was chosen to be the minimum size of a mini-halo \citep[100 kpc at the cluster redshift, as per the definition adopted from][]{2017ApJ...841...71G} and specified as a hard cut in the visibility space using the \texttt{--Selection-UVRangeKm parameter in DDF}. All sources expected to be subtracted were included in the mask derived from the initial HR deconvolution. We used a generic 4$\sigma$ threshold for all clusters to mask the field sources and manually inserted masked pixels that corresponded to confirmed point-like sources in the immediate vicinity of the mini-halo that were not picked up by the generic mask; a generic threshold any tighter than that stated above caused significant artefacts to be included in the mask. It is expected that any emission too faint to be captured by the HR mask is left unsubtracted. The deep HR model derived from the deeper deconvolution needs to be (re)-predicted in DDF with \texttt{--Output-Mode Predict} while specifying the same uv-cut that will be used in the imaging of the diffuse emission. This ensures that the shorter baselines that were previously cut are included into the calculation of the subtraction model, i.e. so that all spatial scales are considered during every step of modelling, subtraction and (re)-imaging. Once this final model was set, we used the \texttt{msutils} method of Stimela \citep{makhathini2018} to perform the subtraction operation. The resulting column was then imaged at \texttt{robust} 0, giving an $\sim$8$\arcsec$ resolution. External masks were thereafter generated using the same thresholds from Step 2 of Section \ref{section:3GC}, ultimately giving the final source-subtracted (SRC-SUB) 2GC/3GC images. 
	
	The basic procedure is summarized below:
	\begin{enumerate}
		\item \textit{Initial deconvolution at HR;}  
		\item \textit{HR masking;} 
		\item \textit{Deeper deconvolution;} 
		\item \textit{Re-predict HR model;}
		\item \textit{Subtraction;} 
		\item \textit{Initial deconvolution at standard resolution;}
		\item \textit{Final deconvolution.}
	\end{enumerate}
	The Aladin Sky Atlas \citep{2000A&AS..143...33B,2014ASPC..485..277B} was used to overlay varies catalogues on the images to identify individual sources when needed. The only differences between the 2GC and 3GC subtraction were the implementation of the kMS solutions and the masking thresholds.
	
	\subsection{Spectral imaging}
	\label{section:spectral_imaging}
	
	There are no spectral index maps available for the mini-halos in our sample, and so we generate them for the first time with this MeerKAT data. We used the sub-band images to analyse both the unsubtracted and SRC-SUB in-band spectral properties of each mini-halo. To do this, we repeated the 3GC and source subtraction procedures from Step 5 of Section \ref{section:3GC} with three equal sub-bands, centred at 998, 1283 and 1569 MHz, and fitting a 3rd-order spectral polynomial. For the \texttt{robust} 0 images, to ensure the same uv-range was imaged across all sub-bands, we applied an inner and outer taper of $\sim$5$\arcmin$ and $\sim$14$\arcsec$, respectively. We then convolved each sub-band image to a common 15$\arcsec$ circular beam, except for the SRC-SUB ACO 1413 images. In this case, significant oversubtraction of the embedded sources occurred in the high sub-band image and so, to salvage a detection, we convolved to 30$\arcsec$. Similarly, for the HR images, to determine in-band spectral properties of the embedded sources, we applied an inner taper corresponding to the uv-cut and an outer taper $\sim$15$\arcsec$. The resulting sub-band images were convolved to a 7$\arcsec$ resolution; except for ACO 1413, which were convolved to a 12$\arcsec$ because of its high northern declination and poorer spatial resolution, particularly in the low-band image. We used \texttt{brats} \citep{10.1093/mnras/stt1526,10.1093/mnras/stv2194} to make the in-band spectral index maps, with a $5\sigma$ cutoff and a 10\% calibration uncertainty.
	
	\subsection{ACO 1795}\label{section:1795}
	
	Strong artefacts centred around the cluster BCG limited image quality and prevented us from detecting any nearby diffuse emission (see Figure \ref{fig:1795} in Appendix \ref{appendix: 1795}). This BCG is known to be particularly strong; $917\pm46$ mJy from its FIRST image \citep{2014ApJ...781....9G} and $975 \pm 97.5$ mJy from our best 2GC and 3GC images (left and right panels of Figure \ref{fig:1795}). Its strong sidelobes could not be sufficiently corrected by the calibration methods used for the rest of the clusters.
	
	To improve the accuracy of the 2GC calibration solutions described in Section \label{section:2GC}, we used QuartiCal (CubiCal's successor, currently in public beta release) to chain together Jones terms during individual rounds of DI self-calibration. In each round, for three rounds, we simultaneously derived and applied a delay solution and (per channel) complex-gain solutions, where the latter was refined down to 60 equal time intervals by the third round. For more controlled cleaning, since we noticed some artefacts appearing in the CARACal model images, we performed manual masking with \texttt{breizorro} similarly to that described in Section \ref{section:3GC}, again finishing on 3$\sigma$. We found significant improvements over the CARACal images but still no diffuse emission could be detected. Furthermore, no improvements were ever found after 3GC. We suspect polarisation leakage effects compounded by the poor uv-sampling distribution and asymmetric PSF due to the target's high northern declination ($+26^{\circ}$) to be the cause of the residual BCG sidelobe modelling errors. The calibration may have improved if standard primary and secondary calibrators were observed together with 3C286. Further investigation is required to determine if synergy between MeerKAT/SKA-Mid and LOFAR is possible for this cluster. We exclude ACO 1795 from all further discussions. Below we describe the results for the remaining clusters. 
	
	\section{Results}
	\label{section:results}
	
	This section presents the mini-halo cluster images and associated statistical and physical properties. We give qualitative statistics and depictions (Table \ref{tab:Table D.1} and Figures \ref{fig:A1413_artifacts}-\ref{fig:MACSJ2140.2-2339_artifacts} in Appendix \ref{appendix: 2vs3gc}) of the improvements in image quality after using the 3GC procedure outlined in Section \ref{section:3GC}. In Figures \ref{fig:A1413_radio_mini_halo}-\ref{fig:MACSJ2140.2-2339_radio_mini_halo}, for each mini-halo, we show: (i) a central cutout of our standard resolution (\texttt{robust} 0, $\sim$8$\arcsec$) images before and after point source subtraction in panels a) and b), (ii) a lower resolution (LR, $15\arcsec$) image of b) in panel c) from which we derive flux density and size measurements, (iii) an in-band spectral index and associated uncertainty map of the unsubtracted mini-halo in panels d) and e). Lastly, we give an in-band integrated spectrum of each SRC-SUB mini-halo, shown in Figure \ref{fig:1D_spectra}. 
	
	\begin{table*}
		\centering
		\caption{Source measurements. Columns: cluster name, source, flux density at 1.28 GHz, average mini-halo diameter (to the nearest kpc), largest linear size (to the nearest kpc), spectral index, k-corrected radio power at 1.4 GHz.\\
			$^{*}$ Sub-band maps convolved to 30$\arcsec$.\\
			$^{**}$ Spectral index of the unresolved source S1+S2.\\
			$^{***}$ A large subtraction uncertainty in the flux density measurements due to the strong AGN causes a large uncertainty in the integrated spectrum.}
		\begin{tabular}{ccccccc}
			\hline
			Cluster  & Source    & \makecell{$S_{\rm 1.28\,GHz}$ \\ (mJy)}  & \makecell{$D_{\rm MH}$ \\ (kpc)} & \makecell{LLS \\ (kpc)}   &  $\alpha$ & \makecell{$P_{\rm 1.4\,GHz}$ \\      ($\times 10^{24}$ W/Hz)}                \\
			\hline
			\multirow{3}{*}{ACO 1413}           & Mini-halo      & $2.05 \pm 0.27$    & 185 & 211 & $1.52 \pm 0.46^{*}$&  $0.104 \pm 0.014$    \\
			& BCG (S1)       & $0.37 \pm 0.05$   & &  & $0.89 \pm 0.26^{**}$ &$0.018 \pm 0.003$   \\           
			& S2             & $2.86 \pm 0.29$  & & & $0.89 \pm 0.26^{**}$ & $0.143 \pm 0.015$  \\
			\\
			\multirow{2}{*}{ACO 3444}            & Mini-halo     & $12.10 \pm 1.71$  & 372 & 412 & $1.53 \pm 0.44$  & $2.40 \pm 0.40$        \\
			& BCG           & $2.25 \pm 0.24$  & & & $0.61 \pm 0.27$  & $0.38 \pm 0.04$       \\          
			\\
			\multirow{2}{*}{MACS J1115.8+0129}   & Mini-halo     & $7.91 \pm 2.59$ & 375 & 499 &  $1.00 \pm 1.10^{***}$ & $3.00 \pm 1.20$        \\
			& BCG           & $8.73 \pm 0.88$ & & & $0.78 \pm 0.26$ & $3.15 \pm 0.40$           \\               
			\\
			\multirow{2}{*}{MACS J2140.2-2339}   & Mini-halo    & $2.61 \pm 0.31$ & 296 & 390 &$1.21 \pm 0.36$  & $0.79 \pm 0.11$           \\
			& BCG         & $1.43 \pm 0.15$ & & & $0.72 \pm 0.31$ & $0.39 \pm 0.05$            \\ 
			\hline
		\end{tabular}
		\label{tab:Table 3}
	\end{table*}
	
	Table \ref{tab:Table D.1} shows that all images improved after 3GC, evident by the reduction in global noise levels and greater DR values. We define $DR^{(1)}$ and $DR^{(2)}$ as the maximum pixel value of the unsubtracted image divided by, respectively, its global image rms and the modulus of its minimum pixel value. For the sample, the overall improvement is shown by the global noise and $DR^{(1)}$ values improving by an average of 7\% each, and the factor of three improvement in the $DR^{(2)}$ values showing specifically that the kMS solutions well characterised the deep negative DI point spread function (PSF) sidelobes. The 3GC HR images experienced a similar improvement in $DR^{(2)}$, but also a slight noise amplification in some facets covering the first negative sidelobe of the primary beam causing the global noise to remain near that of the 2GC images. Further investigation into what causes this effect is needed; nevertheless, our science case was not affected. A handful of weak PyBDSF-detected \citep{2015ascl.soft02007M} sources (< 5 mJy) in the 2GC fields experienced flux suppression, decreasing by up to an order of magnitude and resulting in an average flux density difference between 2GC and 3GC of 8\% - further investigation into these outliers is needed. Nevertheless, this difference did not affect our science case as the measured 2GC and 3GC mini-halo flux densities were well consistent within the measured uncertainties. We also note that the 2GC image of MACS J2140.2-2339 was largely absent of any artefacts; therefore, 3GC was not expected to significantly improve this field.
	
	%In fields where no minor and/or major artefacts were present in the 2GC image (MACS J2140.2-2339), the DD calibration could not significantly improve the image quality. The subtraction decreased the maxima values by an average of 96\%, but also degraded the $DR^{min}$ values by 895\% which is mainly due to the excessive minima that the subtraction produces. However, since the rms values increased by only 4\% and the beam sizes were not affected, the oversubtraction contributed minimally to the overall image noise and we judge the subtracted images to still be of sufficient quality for our science case. The DD SRC-SUB images have maxima peaks less than or equal to, and minima peaks greater than or equal to, their DI counterparts suggesting that the DD calibration improved the extent to which the subtracted sources were modelled, but since the DI and DD HR images are comparable, the improvement in subtraction most likely stems from the deeper masks achievable in the \texttt{robust} 0 DD images.
	
	The images at the native $\sim$8$\arcsec$ resolution show the detailed structure of the mini-halos; however, to enhance the surface brightness sensitivity we convolved these to a low-resolution (LR) $15\arcsec$ circular beam, as shown in panel c) of Figures \ref{fig:A1413_radio_mini_halo}-\ref{fig:MACSJ2140.2-2339_radio_mini_halo}. From these LR images, we measure the flux density, radio power and size within the $3\sigma$ contours (see Table \ref{3}). We followed \cite{2014ApJ...781....9G} in calculating the uncertainty on the flux density measurements as the quadrature sum of the systematic, statistical and subtraction uncertainties. The latter is given by:
	\begin{equation}\label{2}
		\sigma_{\rm sub}^{2} = \sum_{\mathclap{s=1}}^{N} \left(I_{\rm MH,\,s} \times N_{\rm beam,\,s}\right)^{2}
	\end{equation}
	where $I_{\rm MH,\,s}$ is the mean residual surface brightness of the mini-halo within the $s^{\rm th}$ source region, and $N_{\rm beam,\,s}$ is the number of beams within that region. Combining these uncertainties gives:
	\begin{equation}\label{1}
		\sigma_{\rm MH} =  \sqrt{ \left(\sigma_{\rm cal} S_{\rm MH}\right)^{2} + \left( rms\sqrt{N_{\rm beam}}\right)^{2} + \sigma_{\rm sub}^{2}} 
	\end{equation}
	where $\sigma_{\rm cal}$ is the percentage calibration uncertainty, $S_{\rm MH}$ is the measured mini-halo flux density, rms is the local noise given in each the caption of each respective image and $N_{\rm beam}$ is the number of beams contained within the mini-halo region. We found an average calibration uncertainty of 10\% across our sample after cross-matching and comparing the 3GC flux densities of the PyBDSF-detected sources > 1 mJy within the primary beam of the MeerKAT images, scaled using a spectral index of 0.7, to those from the NVSS catalogue \citep{1998AJ....115.1693C}. Since mini-halos usually have non-spherical morphologies, we followed \cite{2007MNRAS.378.1565C} in measuring the average diameter as
	\begin{equation}\label{3}
		D_{\rm MH} = \sqrt{D_{\rm min} \times D_{\rm max}}
	\end{equation}
	where $D_{\rm min}$ and $D_{\rm max}$ are the minimum and maximum diameters of the $3\sigma$ contours. Note that $D_{\rm max}$ is the same as the largest linear size (LLS). We measure the BCG flux and associated uncertainty in a similar manner but derived from the HR images, and are listed in Table \ref{3} as well. 
	
	During the spectral analysis, we measured the flux densities within the $5\sigma$ contours of each sub-band image. We fit these to a power-law least squares regression across the central frequencies to produce an in-band integrated spectrum for each SRC-SUB mini-halo and BCG - Table \ref{tab:Table 3} shows the fitted slopes. Note that we choose to display the unsubtracted spectral index maps while we use the SRC-SUB integrated spectrum to describe the average mini-halo spectral index. This is because characterizing the spectral effects of our subtraction method on the residual diffuse emission is beyond the scope of this work. The resulting large spectral uncertainties (which are absolute in nature) limits our discussion on the science extracted from the spectral analyses.
	
	\section{Discussion}
	\label{section:discussion}
	
	The 3GC procedure discussed in Section \ref{section:3GC} improved image quality in each case by reducing artefacts and thereby lowering the background rms noise level, allowing for more significant emission to be detected. We then described the 2GC/3GC visibility-plane point source subtraction and applied it to our cluster sample in Section \ref{section:src_sub} to image the central mini-halos without bias from the surface brightness of the BCG. Section \ref{section:results} gave the resulting statistics that evaluated the reduction and the physical quantities describing the mini-halo systems. This section discusses these result in the context of previous literature findings. We used SIMBAD\footnote{\url{http://simbad.cds.unistra.fr/simbad/}} to identify individual sources of interest unless otherwise stated.
	
\begin{figure*}
		\centering
		\includegraphics[width=\textwidth]{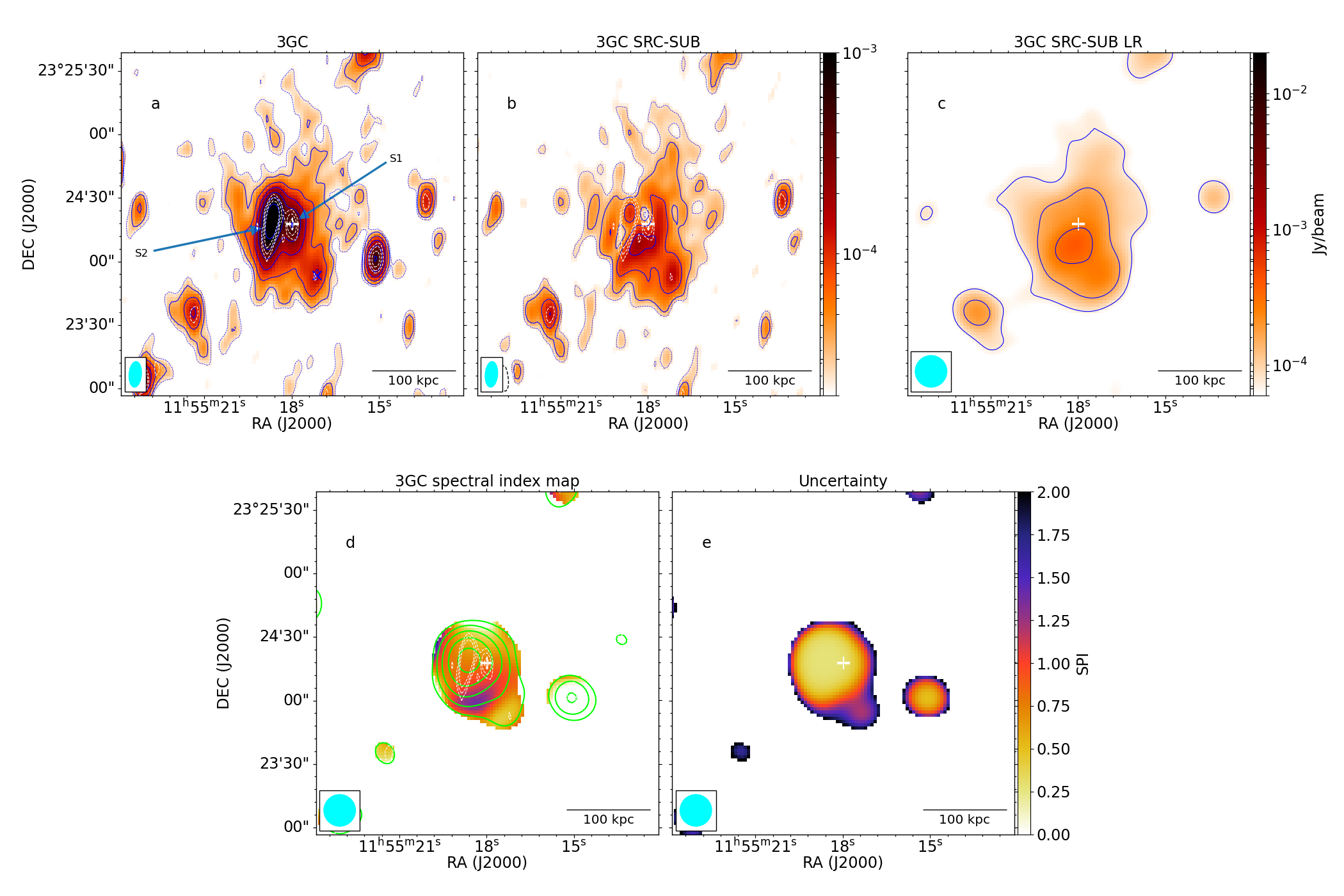}
		\caption{ACO 1413 3GC mini-halo. a) beam size $(12.2\arcsec, 5.9\arcsec, -4.5^{\circ})$, local rms $1\sigma = $ 11.2 $\mu$Jy/beam, S1 marks the BCG, S2 and a small blue cross indicate other projected galaxies. b) SRC-SUB image, beam size and color-scale same as a), local rms $1\sigma = $ 12.2 $\mu$Jy/beam. c) SRC-SUB 15$\arcsec$ LR image, local rms $1\sigma = 25.1$ $\mu$Jy/beam. d) unsubtracted spectral index map and e) uncertainty map. The beam is shown in cyan in the bottom left corner of the images. Dashed blue contours show the $2\sigma$ level and the solid blue contours start at $3\sigma$ and increase by a factor of 2. Dashed black contours show the $-3\sigma$ level. White contours show the emission from the HR image, which has beam size $(7.6\arcsec, 3.1\arcsec, -2.6^{\circ})$ and local rms $1\sigma = $ 23.0 $\mu$Jy/beam. Green contours in d) show the surface brightness of the SRC-SUB mid-band image and start at $5\sigma$ and increase by a factor of 2, where $1\sigma = 25.7$ $\mu$Jy/beam. The white plus indicates the BCG position.}
		\label{fig:A1413_radio_mini_halo}
	\end{figure*}

\begin{figure*}
		\centering
		\includegraphics[width=\textwidth]{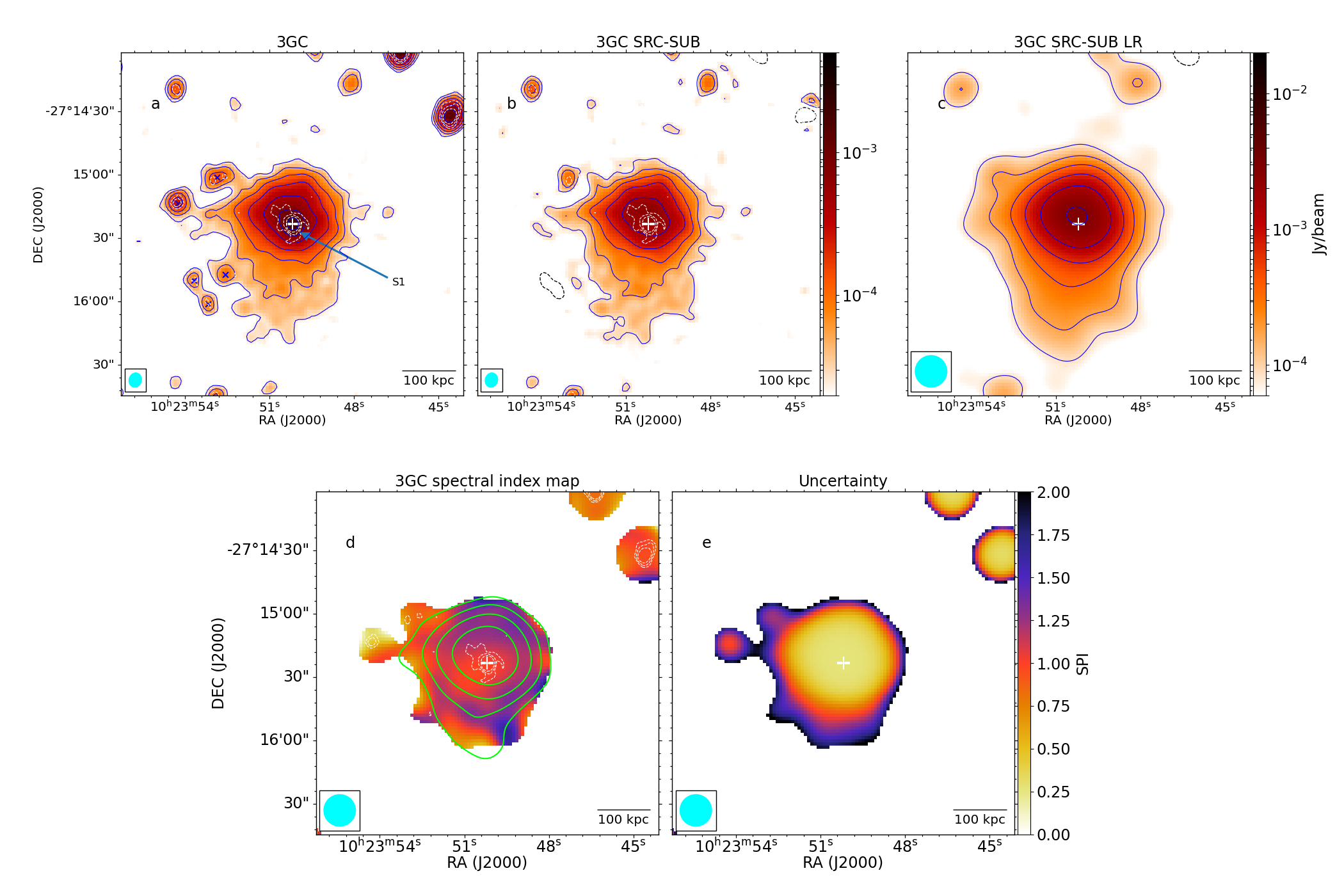}
		\caption{ACO 3444 3GC mini-halo. a) beam size $(6.8\arcsec, 5.9\arcsec, -16.0^{\circ})$, local rms $1\sigma = 8.5$ $\mu$Jy/beam, S1 marks the BCG, blue crosses indicate individual sources that were manually masked at HR. b) SRC-SUB image, beam size and color-scale same as a), local rms $1\sigma = 9.3$ $\mu$Jy/beam. c) SRC-SUB 15$\arcsec$ LR image, local rms $1\sigma = 27.8$ $\mu$Jy/beam. d) unsubtracted spectral index map and e) uncertainty map. The beam is shown in cyan in the bottom left corner of the images. Blue contours start at $3\sigma$ and increase by a factor of 2. Dashed black contours show the $3\sigma$ level. White contours show the emission from the HR image, which has beam size $(3.9\arcsec, 3.2\arcsec, -9.6^{\circ})$ and local rms $1\sigma = $ 20.9 $\mu$Jy/beam. Green contours in d) show the surface brightness of the SRC-SUB mid-band image and start at $5\sigma$ and increase by a factor of 2, where $1\sigma = 34.6$ $\mu$Jy/beam. The white plus indicates the BCG position.}
		\label{fig:A3444_radio_mini_halo}
	\end{figure*}

\begin{figure*}
		\centering
		\includegraphics[width=\textwidth]{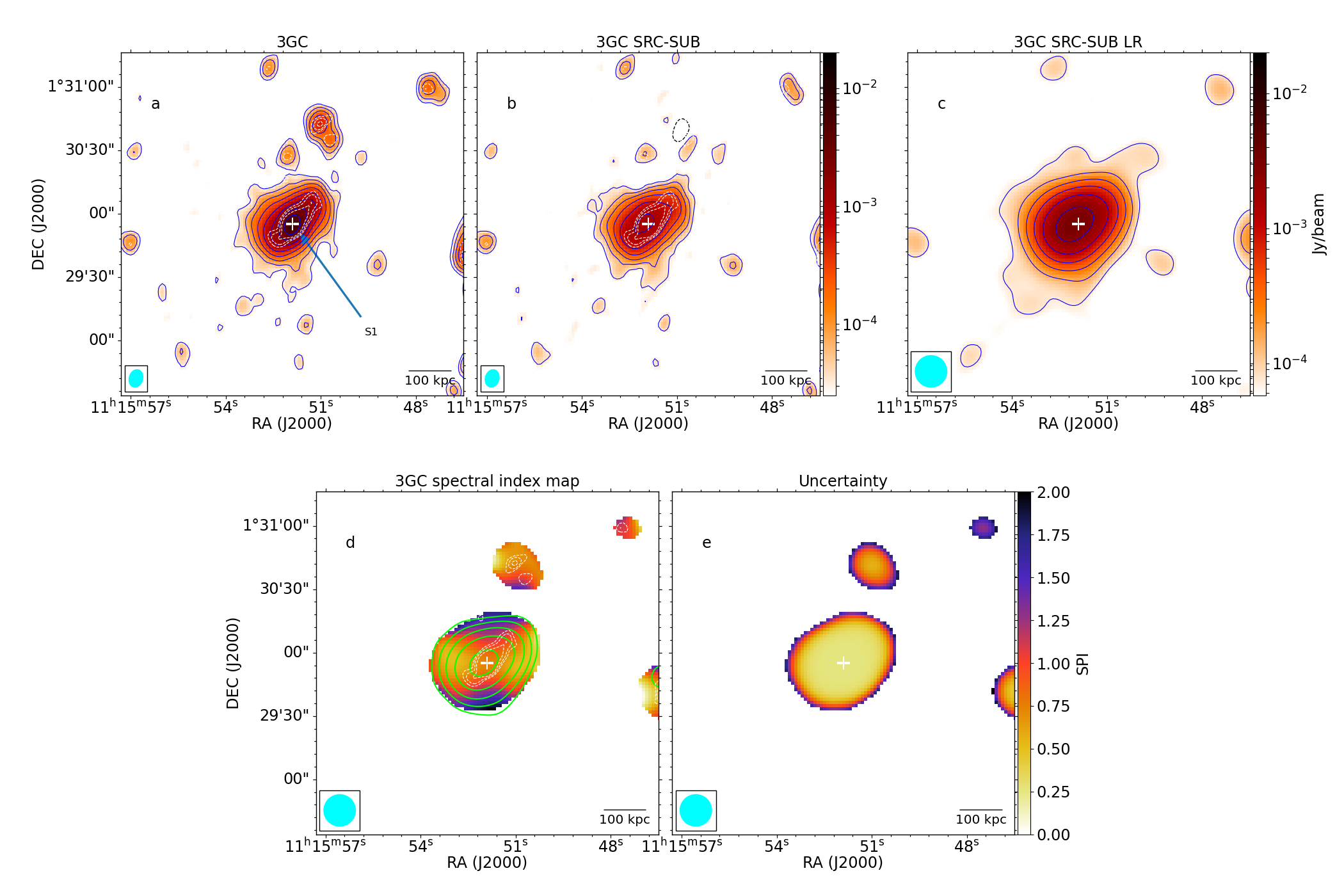}
		\caption{MACS J1115.8+0129 3GC mini-halo. a) beam size $(8.5\arcsec, 6.5\arcsec, -16.2^{\circ})$, local rms $1\sigma = 10.2$ $\mu$Jy/beam, S1 marks the BCG. b) SRC-SUB image, beam size same and color-scale as a), local rms $1\sigma = 10.9$ $\mu$Jy/beam. c) SRC-SUB 15$\arcsec$ LR image, local rms $1\sigma = 24.6$ $\mu$Jy/beam. d) unsubtracted spectral index map and e) uncertainty map. The beam is shown in cyan in the bottom left corner of the images. Blue contours start at $+3\sigma$ and increase by a factor of 2. Dashed black contours show the $3\sigma$ level. White contours show the sources from the HR image, which has beam size $(4.5\arcsec, 4.0\arcsec, -39.3^{\circ})$ and local rms $1\sigma = $ 28.6 $\mu$Jy/beam. Green contours in d) show the surface brightness of the SRC-SUB mid-band image and start at $5\sigma$ and increase by a factor of 2, where $1\sigma = 31.8$ $\mu$Jy/beam. The white plus indicates the BCG position.}
		\label{fig:MACSJ1115.8+0129_radio_mini_halo}
	\end{figure*}

	It is difficult to quantify the effectiveness of a subtraction procedure. In our case, some strong sources were oversubtracted while many more fainter sources were undersubtracted, seen in Figures \ref{fig:A1413_radio_mini_halo}-\ref{fig:MACSJ2140.2-2339_radio_mini_halo} by the residual positive and negative surface brightness regions. Overall, the subtraction induced minimal error, increasing the noise of the SRC-SUB images by an average of 4\%. We notice that for most of our sample, before and after point source subtraction, the mini-halo orientation roughly follows that of the BCG. The role of the AGN in mini-halo formation is currently still uncertain; however, strong correlations between mini-halo radio power and BCG/X-ray cavity power suggest a clear physical connection \citep{2020MNRAS.499.2934R}. Further high resolution and sensitive observations, similar to those presented in \citet{2022A&A...659A..20I,2023MNRAS.520L...1P,2023MNRAS.519..767B}, are required to gain a deeper understanding of the connection between the embedded radio galaxy and mini-halo and their orientation as we found in our study.
	%This could be the result of the fundamental role that the BCG plays in the production of the mini-halo emission, possibly being the source of a population of seed relativistic electrons and/or protons which are then (re)-accelerated and (re)-distributed by turbulence and/or hadronic collisions \citep[e.g., ][]{2014IJMPD..2330007B}. Since diffusion and other transport mechanisms can only spread relativistic electrons to a maximum of $\sim$50 kpc within their radiative lifetime \citep{2017ApJ...841...71G} and the mini-halos extend well beyond this limit, we find it unlikely that the above connection is caused by a failure in the subtraction to properly disentangle the BCG and mini-halo emissions, but do not completely dismiss this possibility.
	
	\subsection{ACO 1413}
	
	\citet{2017ApJ...841...71G} classified ACO 1413 as a (modest) non-cool-core cluster using \textit{Chandra} data (Obs. ID 5002). \citet{2009A&A...499..371G} first showed the presence of a candidate mini-halo in ACO 1413 using VLA 1.4 GHz C-configuration data (maximum baseline 3.4 km with L-band resolution 14$\arcsec$), making it the first and only non-cool-core cluster to host a candidate mini-halo. \citet{2019A&A...622A..24S} later used a 144 MHz LOFAR observation to confirm its detection.
	
	Our MeerKAT wide-FoV images of ACO 1413, shown in Figure \ref{fig:A1413_artifacts} of Appendix \ref{appendix: 2vs3gc}, give a depiction of the improvement from 2GC to 3GC imaging. The DI image displays a significant snowflake-like artefact around a bright double radio source located at (RA, DEC) = (11:52:26.5, 23:13:47.3) and (11:52:26.2, 23:12:54.6), designated as NVSS J115226+231347 and NVSS J115226+231255, with flux densities measured from the final 3GC intrinsic images $\sim$260 mJy and $\sim$175 mJy, respectively. The sources are located at a distance of $\sim$40$\arcmin$ from the phase centre (just outside the main lobe of the primary beam) and create significant artefacts affecting the image quality. These artefacts are dramatically reduced after 3GC, with much of the snowflake spikes going down to noise level, drastically improving image quality. The global rms noise improved by 11\% from 8.3 $\mu$Jy/beam to 7.4 $\mu$Jy/beam. To depict how valuable 3GC can be if a source of interest is affected by such artefacts, we show a zoom of an FR-II galaxy located $\sim$33$\arcmin$ from the phase centre in the insets of Figure \ref{fig:A1413_artifacts}. In this comparison, the 2GC image displays a snowflake-spike piercing the radio lobes and nucleus; however, the 3GC image shows these spikes disappeared, giving a clear depiction of the source morphology. The local rms near this source improves by just under a factor of two, from 22 $\mu$Jy/beam to 12 $\mu$Jy/beam after 3GC.
	
	Panels a) and b) of Figure \ref{fig:A1413_radio_mini_halo} show a central cutout comparing the 3GC vs 3GC SRC-SUB images for this cluster. \citet{2019A&A...622A..24S}, in their Figure 6, first noted that there are in fact two sources embedded within this mini-halo, and these are clearly visible in this image; S1 being the BCG with optical counterpart MCG+04-28-097 and S2 being the radio source FIRST J115518.6+232422. A localised peak south-west of these sources is also visible, marked with a small blue cross in Figure \ref{fig:A1413_radio_mini_halo}a, identified as galaxy 2MASS J11551712+2323527 and is blended with the mini-halo. The morphology is similar to that shown in the unsubtracted image of \citet{2019A&A...622A..24S}. The result of the subtraction, Figure \ref{fig:A1413_radio_mini_halo}b, shows only the contribution of diffuse emission in the field. The mini-halo is visible in this image and is shown to have an irregular morphology. It is extended from north to south, similar to the BCG orientation, with the north having fainter filamentary-like emission that is only detected at the $2\sigma$ level.
	
	The 3GC SRC-SUB LR image with circular beam size of 15$\arcsec$ is shown in pancel c) of Figure \ref{fig:A1413_radio_mini_halo}. Once smoothed, the morphology is comparable to the corresponding SRC-SUB image in \citet{2019A&A...622A..24S}, even though our image is at a slightly higher resolution. From this image, we determine a flux density of $S_{\rm 1.28\,GHz} = 2.05 \pm 0.27$ mJy for this mini-halo, and an average diameter of 185 kpc with LLS of 211 kpc. These values are consistent with those presented in \citet{2009A&A...499..371G} and \citet{2019A&A...622A..24S} when scaled with the measured in-band spectral index of $1.52 \pm 0.46$ (see below). Additionally, the k-corrected radio power scaled to 1.4 GHz is $P_{\rm 1.4\,GHz} = (0.104 \pm 0.014) \times 10^{24}$ W/Hz. This is the faintest mini-halo in our sample. 
	
	The spectral analysis of this mini-halo is limited in both sensitivity and resolution due to the cluster's high declination and inconsistent subtraction across the sub-bands. The high-band SRC-SUB image at native resolution showed at the location of source S2 a $\sim$50 kpc region of negative surface brightness where the low- and mid-band images had non-negative brightness. To salvage a measurement of the in-band spectrum, we convolve the subtracted images to 30$\arcsec$ and compute the flux density inside the $3\sigma$ contours. We ultimately obtain an average spectral index of $\alpha^{\rm 1.5\,GHz}_{\rm 1\,GHz} = 1.52 \pm 0.46$, which is consistent with the value of 1.3 given in \citet{2019A&A...622A..24S}. When imaging the HR sub-band images, the low-band image had a very poor resolution and so could not reliably measure the index of each embedded source, S1 and S2. These sources blended to give one unresolved source at a 12$\arcsec$ resolution and thus the spectral index quoted in Table \ref{tab:Table 3} of $\alpha^{\rm 1.5\,GHz}_{\rm 1\,GHz} = 0.89 \pm 0.26$ is of this combined source. The spectral index and uncertainty maps of the unsubtracted mini-halo, which could remain at a 15$\arcsec$ resolution, are shown in panels d) and e) of Figure \ref{fig:A1413_radio_mini_halo}. Sources S1 and S2 obscure the view of the underlying mini-halo spectral distribution, which is evident for all our spectral index maps. Nevertheless, the spectral index of the entire radio core region is reliable within a central $\sim$100 kpc diameter and, outside the embedded source regions, a possible gradient from north to south is visible, with a flatter spectrum of $\sim$0.4 in the north and a steeper spectrum of $\sim$1.3 in the south. This southern region may be a glimpse of the underlying mini-halo as it is not contaminated by any embedded sources and is roughly consistent with the integrated spectrum. The blended infrared source mentioned above causes the flatter spectrum visible in the south-west.
	
	\subsection{ACO 3444}
	
	\citet{2007A&A...463..937V} reported a candidate mini-halo in ACO 3444 based on a 610 MHz GMRT observation. \citet{2017ApJ...841...71G,2019ApJ...880...70G} provided confirmation through the analysis of archival VLA 1.4 GHz DnC+BnA configuration data, with images restored to 5$\arcsec$ and 11$\arcsec$.
	
	\begin{figure*}
		\centering
		\includegraphics[width=\textwidth]{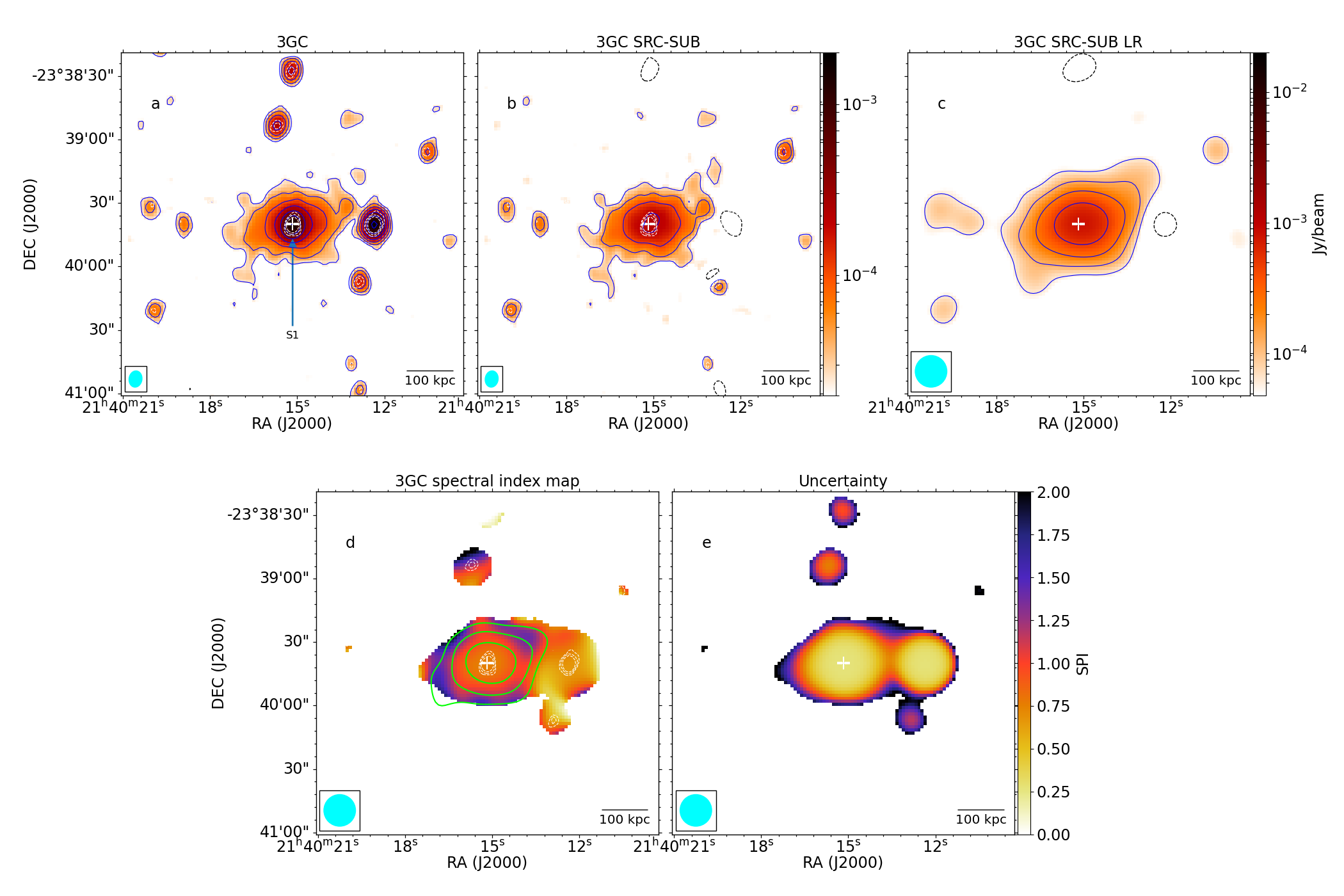}
		\caption{MACS J2140.2-2339 3GC mini-halo. a) beam size $(7.8\arcsec, 6.2\arcsec, -8.6^{\circ})$, local rms $1\sigma = 8.0$ $\mu$Jy/beam, S1 marks the BCG. b) SRC-SUB image, beam size same and color-scale as a), local rms $1\sigma = 8.7$ $\mu$Jy/beam. c) SRC-SUB 15$\arcsec$ LR image, local rms $1\sigma = 20.4$ $\mu$Jy/beam. d) unsubtracted spectral index map and e) uncertainty map. The beam is shown in cyan in the bottom left corner of the images. Blue contours start at $+3\sigma$ and increase by a factor of 2. Dashed black contours show the $3\sigma$ level. White contours show the sources from the HR image, which has beam size $(4.3\arcsec, 3.2\arcsec, -7.1^{\circ})$ and local rms $1\sigma = $ 20.0 $\mu$Jy/beam. Green contours in d) show the surface brightness of the SRC-SUB mid-band image and start at $5\sigma$ and increase by a factor of 2, where $1\sigma = 22.8$ $\mu$Jy/beam. The white plus indicates the BCG position.}
		\label{fig:MACSJ2140.2-2339_radio_mini_halo}
	\end{figure*}
	
	Figure \ref{fig:A3444_artifacts} shows a comparison between the 2GC and 3GC images for this cluster. Significant concentric artefacts that affect the entire north-central region of the image originate from a set of three bright point-like sources located $\sim$30$\arcmin$ from the phase centre. These sources are positioned from left to right as seen in Figure \ref{fig:A3444_artifacts} at (RA, DEC) = (10:22:29.4, -26:47:34.5), (10:23:27.1, -26:50:09.7), (10:24:07.4, -26:44:17.5), designated as WISEA J102407.38-264417.8 (NED), NVSS J102327-265009 and NVSS J102229-264734, respectively, and have a maximum flux density measured from the 3GC intrinsic image to be $\sim$270 mJy. The PSF sidelobes and further artefacts have been reduced by the kMS gain solutions and are now almost at the level of the background noise. The global noise rms improved by 8\% from 6.4 $\mu$Jy/beam to 5.9 $\mu$Jy/beam. The inset of Figure \ref{fig:A3444_artifacts} shows a zoom of the third contaminating source mentioned above; its local rms improved by just over a factor of two, from $\sim$52 $\mu$Jy/beam to $\sim$24 $\mu$Jy/beam.
	
	Figures \ref{fig:A3444_radio_mini_halo}a, b show the comparison between the 3GC vs 3GC SRC-SUB images. The embedded BCG (marked S1, optical counterpart 2MASX J10235019-2715232) and mini-halo are clearly visible. Also visible are five compact sources on the eastern edge of the mini-halo that obscure its boundary, identified from NED to be individual infrared sources and indicated in Figure \ref{fig:A3444_radio_mini_halo}a by small blue crosses. These sources had to be manually added to the HR mask. To directly compare our MeerKAT SRC-SUB image to the VLA image shown in Figure 5b of \citet{2019ApJ...880...70G}, which had twice as much on-target exposure time, we restored it to the same 11$\arcsec$ beam size and found that our image has a local rms 2.6 times deeper and shows the mini-halo extended further south by $\sim$100 kpc. This detection shows the ability of the MeerKAT to detect, with shorter tracks, more low surface brightness L-band emission than what was achieved with the VLA. The extension is partially seen in Figure A4 of \citet{2007A&A...463..937V}, suggesting that it may have a particularly steep spectrum.
	
	Figure \ref{fig:A3444_radio_mini_halo}c shows the LR 3GC SRC-SUB image. Residual diffuse emission, coincident with the northernmost infrared source, blends with the mini-halo but its contribution to the determined size and flux density is negligible. We found a flux density of $S_{\rm 1.28\,GHz} = 12.10 \pm 1.71$, and an average diameter of 372 kpc with LLS of 412 kpc. The flux density is consistent with that presented in \citet{2019ApJ...880...70G}; however, the extension adds another $\sim$130 kpc onto the reported diameter and $\sim$60 kpc onto the \citet{2007A&A...463..937V} LLS. Even though a larger size is observed, a similar flux density is expected because of the extension's low surface brightness. Using the measured in-band spectral index (see below), the k-corrected radio power is $P_{\rm 1.4\,GHz} = (2.40 \pm 0.40) \times 10^{24}$ W/Hz, consistent with \citet{2019ApJ...880...70G}. However, the BCG flux density is significantly higher than that reported in \citet{2019ApJ...880...70G}.
	
	In the spectral analysis of ACO 3444, the southern extension progressively recedes in the higher parts of the band - also suggesting a steep spectrum. The remaining SRC-SUB mini-halo gives a spectral index $\alpha^{\rm 1.5\,GHz}_{\rm 1\,GHz} = 1.53 \pm 0.44$. The BCG has $\alpha^{\rm 1.5\,GHz}_{\rm 1\,GHz} = 0.61 \pm 0.27$. The spectral index maps of the unsubtracted mini-halo are shown in panels d) and e) of Figure \ref{fig:A3444_radio_mini_halo}, where much of the southern extension is undetected. The eastern edge is flatter due to the contaminating infrared sources. No preferential gradient is evident as much of the region outside the BCG has an index consistent with the integrated spectrum.
	
	\subsection{MACS J1115.8+0129}
	
	\citet{2020A&A...640A.108G} reported the presence of a mini-halo in MACS J1115.8+0129 through a short JVLA 1.5 GHz D-configuration observation (maximum baseline 1.03 km with L-band resolution 46$\arcsec$). In this lower-resolution data, the central source blended with the diffuse emission. Hence, the properties of any diffuse structures could not be determined and the authors simply give the total flux density of the entire radio core region (compact+diffuse emission). They compare this total flux density to that of the same region's FIRST counterpart and cite the $\sim$7 mJy difference as evidence for a radio mini-halo.
	
	Figure \ref{fig:MACSJ1115.8+0129_artifacts} shows a comparison between the 2GC and 3GC images for MACS J1115.8+0129. No major artefacts are present in the 2GC image but several sources inside the primary beam cause some minor artefacts. One such source identified as GB6 B1114+0136, located $\sim$22$\arcmin$ south-east of the phase centre, is highlighted by the inset and the comparison shows the reduction in artefacts after 3GC. Its local rms value improves by just under a factor of two, from 22 $\mu$Jy/beam to 12 $\mu$Jy/beam. The artefacts of a few other sources similarly reduce (except for a source $\sim$27$\arcmin$ south of the phase centre), such that the global noise level improves by 7\% from 8.2 $\mu$Jy/beam to 7.6 $\mu$Jy/beam.
	
	\begin{figure}
		\centering
		\includegraphics[width=\columnwidth]{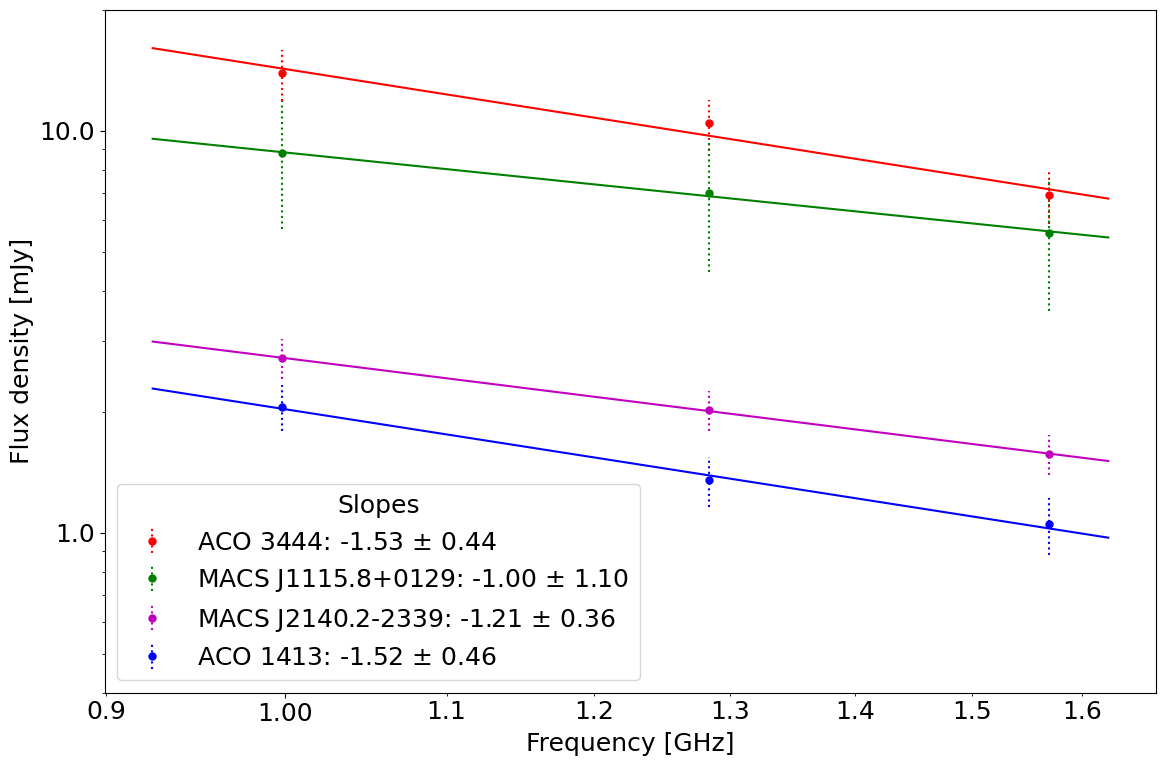}
		\caption{3GC SRC-SUB mini-halo integrated spectra. All sub-band images were convolved to 15$\arcsec$ and flux densities measured within a $5\sigma$ contour, except for ACO 1413 which were convolved to 30$\arcsec$ and measured within $3\sigma$.}
		\label{fig:1D_spectra}
	\end{figure}
	
	Figures \ref{fig:MACSJ1115.8+0129_radio_mini_halo}a, b show the 3GC vs 3GC SRC-SUB images for this cluster. The BCG (optical counterpart SDSS J111551.90+012955.0) is shown to have radio lobes orientated in the north-west/south-east direction, which are removed in the SRC-SUB image showing the residual diffuse emission. After such subtraction, the mini-halo is shown to have a similar orientation. We are limited by sensitivity to confirm the full extent of its morphology; however, we note that there appears to be a hint of a radio arm extending from the south, seemingly misaligned with the rest of the mini-halo.
	%; additionally, connection between AGN and diffuse emission
	%Hence, there may be a morphological connection to the AGN, contrary to the adopted definition. Further investigation into the exact connection between AGN and diffuse emission is required. However, for now, because the overall morphology of the diffuse emission remains unchanged after subtraction, we still consider this source a mini-halo. 
	
	Figure \ref{fig:MACSJ1115.8+0129_radio_mini_halo}c shows the 3GC SRC-SUB image at a 15$\arcsec$ resolution. The convolution causes some nearby faint emission to blend with the mini-halo along the direction of its LLS, thus the sizes for this mini-halo quoted in Table \ref{tab:Table 3} may be slightly overestimated. The strong embedded AGN results in large measurement uncertainties for the mini-halo flux density, integrated spectrum and radio power, which are dominated by the subtraction uncertainty. We determine a flux density of $S_{\rm 1.28\,GHz} = 7.91 \pm 2.59$, and an average diameter of 375 kpc with LLS of 499 kpc. The flux density of the mini-halo is consistent with the missing flux cited by \citet{2020A&A...640A.108G}, and the BCG flux density is consistent with their FIRST measurement. Using the measured in-band spectral index (below), the k-corrected radio power of the mini-halo is $P_{\rm 1.4\,GHz} = (3.00 \pm 1.20) \times 10^{24}$ W/Hz, making it the most powerful in our sample. 
	
	The integrated spectra gives $\alpha^{\rm 1.5\,GHz}_{\rm 1\,GHz} = 1.0 \pm 1.1$ for the mini-halo and $\alpha^{\rm 1.5\,GHz}_{\rm 1\,GHz} = 0.78 \pm 0.26$ for the BCG. The spectral index and its uncertainty map are shown in panels d) and e) of Figure \ref{fig:MACSJ1115.8+0129_radio_mini_halo} and shows a very interesting spectral distribution; a significant spectral flattening is visible horizontally across the mini-halo and reaches its minimum directly east and west of the AGN core, at a spectral index of $\sim$0.5. The steepest regions are directly north and south of the AGN core, with spectral index $\sim$1.5. 
	%Interestingly, the flattest spectral indices are not spatially coincident with the jets, suggesting that a complex dynamical process is at play.
	
	\subsection{MACS J2140.2-2339}
	
	We present a new mini-halo detection in the galaxy cluster MACS J2140.2-2339. This is an X-ray luminous and massive cool-core cluster at redshift z = 0.313 \citep{2010MNRAS.407...83E}, with $L_{X, 500} = 11.1\times10^{44}$ erg s$^{-1}$ and $M_{500} = 4.7\times10^{14}M_{\odot}$, making it one of the lowest mass clusters to host a confirmed mini-halo. \citet{2018ApJ...853..100Y} produced a high-resolution image of the cluster BCG using a 1.5 GHz JVLA observation in A-configuration (maximum baseline 36.4 km with L-band resolution 1.3$\arcsec$), measured a flux density of $1.39\pm0.03$ mJy and quoted a $3.8\pm0.5$ mJy flux density from the NVSS counterpart. Similar to MACS J1115.8+0129, \citet{2020A&A...640A.108G} observed this cluster with a short JVLA 1.5 GHz observation in D-configuration; however, a useful image could not be produced due to strong RFI. They cite the difference in flux density between the NVSS and JVLA A-configuration measurements of \citet{2018ApJ...853..100Y} as evidence for possible extended radio emission around the BCG. The MeerKAT's southern location, high sensitivity and array layout make it the ideal telescope to observe this source and disentangle the compact and possible diffuse radio emission.
	
	Figure \ref{fig:MACSJ2140.2-2339_artifacts} shows a comparison between the 2GC and 3GC images. Similar to MACS J1115.8+0129, no strong sources causing any significant artefacts are present in the map. Further, there are no major DD errors present, except for minor errors around a source identified as PMN J2142-2329 and highlighted by the inset. The 3GC procedure reduced these artefacts and improved the local rms near this source by just under a factor of 1.75, from 33 $\mu$Jy/beam to 19 $\mu$Jy/beam. Otherwise, no significant improvements could be made over the 2GC image.     
	
	Figures \ref{fig:MACSJ2140.2-2339_radio_mini_halo}a, b compare the 3GC vs 3GC SRC-SUB images. Diffuse emission is clearly evident around the BCG (optical counterpart 2MASX J21401517-2339398), and satisfies all the criteria defining a radio mini-halo set out in \citet{2017ApJ...841...71G}. Thus, we classify this source as a new mini-halo detection. The BCG and high-redshift galaxy on the western edge \citep[CLASH MS2137-2353 830, z = 5.9, ][]{2014ApJ...792...76B} are removed after subtraction, providing an uncontaminated view of the mini-halo, which seems to be smoothly orientated in the east-west direction with no AGN jets influencing its morphology. There seems to be a hint of a pair of radio arms that are symmetrically opposite the core.
	
	Figure \ref{fig:MACSJ2140.2-2339_radio_mini_halo}c shows the $15\arcsec$ LR 3GC SRC-SUB image. We determine a flux density of $S_{\rm 1.28\,GHz} = 2.61 \pm 0.31$, consistent with the difference in measurements from \citet{2018ApJ...853..100Y}. We determine an average diameter of 296 kpc with LLS of 390 kpc. The k-corrected mini-halo radio power is $P_{\rm 1.4\,GHz} = (0.79 \pm 0.11) \times 10^{24}$ W/Hz. The measured BCG flux density is consistent with that given in \citet{2018ApJ...853..100Y}.
	
	The integrated spectra gives $\alpha^{\rm 1.5\,GHz}_{\rm 1\,GHz} = 1.21 \pm 0.36$ for the mini-halo, and $\alpha^{\rm 1.5\,GHz}_{\rm 1\,GHz} = 0.72 \pm 0.31$ for the BCG. The spectral index maps of the unsubtracted data are shown in panels d) and e) of Figure \ref{fig:MACSJ2140.2-2339_radio_mini_halo}. The BCG and western infrared source obscure much of the underlying mini-halo spectral distribution. However, we can see that the spectral index monotonically increases radially outwards from $\sim$0.75 to $\sim$1.50.

	\section{Summary and Conclusions}
	\label{section:summary}
	
	In this paper, we have described the application of 3GC to MeerKAT L-band continuum data for a sample of five relaxed galaxy clusters. We used CARACal for DI calibration and DDF+kMS for DD calibration, and presented Stokes I continuum and spectral index maps of the central radio mini-halos. Our 3GC procedure drastically improved image quality, reducing the global image noise levels and improving the DR$^{(1)}$ values by 7\%, DR$^{(2)}$ by almost a factor of three and the local noise around severe artefact-affected sources by a factor of two. We presented the new detection of a mini-halo in MACS J2140.2-2339, and a $\sim$100 kpc southern extension to the ACO 3444 mini-halo which was not present in previous VLA L-band observations. We were unable to image the candidate mini-halo in ACO 1795 due to significant image artefacts caused by its strong BCG. The point source subtraction procedure allowed for the underlying mini-halos to be viewed but simultaneously increased the global image noise by 4\%. We presented the flux density, size and in-band spectral analysis of these mini-halos, noting that the spectral index maps are generated from the unsubtracted mini-halo images. The described calibration procedures of our MeerKAT wide-band data will be useful for future studies of extended diffuse radio sources. 
	
	The spectral analysis presented here would greatly be furthered by follow-up MeerKAT UHF-band observations. Additionally, a multi-wavelength study of this sample in which the radio and X-ray core characteristics are compared would further the understanding of various correlations between the thermal and non-thermal properties of the host galaxy clusters in the context of the broader population of mini-halos and possible production mechanisms. 
	
	\section*{Acknowledgements}
	
	The MeerKAT telescope is operated by the South African Radio Astronomy Observatory, which is a facility of the National Research Foundation, an agency of the Department of Science and Innovation. The financial assistance of the South African Radio Astronomy Observatory (SARAO) towards this research is hereby acknowledged (www.sarao.ac.za). V. P. acknowledges the South African Research Chairs Initiative of the Department of Science and Technology and National Research Foundation who supported this research. V.P. also acknowledges the financial assistance of the South African Radio Astronomy Observatory (SARAO) towards this research. Oleg Smirnov's research is supported by the South African Research Chairs Initiative of the Department of Science and Technology and National Research Foundation. S. G. acknowledges that the basic research in radio astronomy at the Naval Research Laboratory is supported by 6.1 Base funding.
	
	\section*{Data Availability}
	
	The raw data underlying this article is available in SARAO archive, at \url{https://archive.sarao.ac.za}. 
	
	
	
	\bibliographystyle{mnras}
	%\bibliography{main_revised} % if your bibtex file is called example.bib

\begin{thebibliography}{}
\makeatletter
\relax
\def\mn@urlcharsother{\let\do\@makeother \do\$\do\&\do\#\do\^\do\_\do\%\do\~}
\def\mn@doi{\begingroup\mn@urlcharsother \@ifnextchar [ {\mn@doi@}
  {\mn@doi@[]}}
\def\mn@doi@[#1]#2{\def\@tempa{#1}\ifx\@tempa\@empty \href
  {http://dx.doi.org/#2} {doi:#2}\else \href {http://dx.doi.org/#2} {#1}\fi
  \endgroup}
\def\mn@eprint#1#2{\mn@eprint@#1:#2::\@nil}
\def\mn@eprint@arXiv#1{\href {http://arxiv.org/abs/#1} {{\tt arXiv:#1}}}
\def\mn@eprint@dblp#1{\href {http://dblp.uni-trier.de/rec/bibtex/#1.xml}
  {dblp:#1}}
\def\mn@eprint@#1:#2:#3:#4\@nil{\def\@tempa {#1}\def\@tempb {#2}\def\@tempc
  {#3}\ifx \@tempc \@empty \let \@tempc \@tempb \let \@tempb \@tempa \fi \ifx
  \@tempb \@empty \def\@tempb {arXiv}\fi \@ifundefined
  {mn@eprint@\@tempb}{\@tempb:\@tempc}{\expandafter \expandafter \csname
  mn@eprint@\@tempb\endcsname \expandafter{\@tempc}}}

\bibitem[\protect\citeauthoryear{{Asad} et~al.,}{{Asad}
  et~al.}{2021}]{2021MNRAS.502.2970A}
{Asad} K.~M.~B.,  et~al., 2021, \mn@doi [\mnras] {10.1093/mnras/stab104}, \href
  {https://ui.adsabs.harvard.edu/abs/2021MNRAS.502.2970A} {502, 2970}

\bibitem[\protect\citeauthoryear{{B{\'e}gin} et~al.,}{{B{\'e}gin}
  et~al.}{2022}]{2022arXiv220201235B}
{B{\'e}gin} T.,  et~al., 2022, arXiv e-prints, \href
  {https://ui.adsabs.harvard.edu/abs/2022arXiv220201235B} {p. arXiv:2202.01235}

\bibitem[\protect\citeauthoryear{{B{\'e}gin} et~al.,}{{B{\'e}gin}
  et~al.}{2023}]{2023MNRAS.519..767B}
{B{\'e}gin} T.,  et~al., 2023, \mn@doi [\mnras] {10.1093/mnras/stac3526}, \href
  {https://ui.adsabs.harvard.edu/abs/2023MNRAS.519..767B} {519, 767}

\bibitem[\protect\citeauthoryear{{Biava} et~al.,}{{Biava}
  et~al.}{2021}]{2021MNRAS.508.3995B}
{Biava} N.,  et~al., 2021, \mn@doi [\mnras] {10.1093/mnras/stab2840}, \href
  {https://ui.adsabs.harvard.edu/abs/2021MNRAS.508.3995B} {508, 3995}

\bibitem[\protect\citeauthoryear{{Boch} \& {Fernique}}{{Boch} \&
  {Fernique}}{2014}]{2014ASPC..485..277B}
{Boch} T.,  {Fernique} P.,  2014, in {Manset} N.,  {Forshay} P.,  eds,
  Astronomical Society of the Pacific Conference Series Vol. 485, Astronomical
  Data Analysis Software and Systems XXIII. p.~277

\bibitem[\protect\citeauthoryear{{Bonnarel} et~al.,}{{Bonnarel}
  et~al.}{2000}]{2000A&AS..143...33B}
{Bonnarel} F.,  et~al., 2000, \mn@doi [\aaps] {10.1051/aas:2000331}, \href
  {https://ui.adsabs.harvard.edu/abs/2000A&AS..143...33B} {143, 33}

\bibitem[\protect\citeauthoryear{{Bradley} et~al.,}{{Bradley}
  et~al.}{2014}]{2014ApJ...792...76B}
{Bradley} L.~D.,  et~al., 2014, \mn@doi [\apj] {10.1088/0004-637X/792/1/76},
  \href {https://ui.adsabs.harvard.edu/abs/2014ApJ...792...76B} {792, 76}

\bibitem[\protect\citeauthoryear{{Briggs}}{{Briggs}}{1995}]{1995AAS...18711202B}
{Briggs} D.~S.,  1995, in American Astronomical Society Meeting Abstracts. p.
  112.02

\bibitem[\protect\citeauthoryear{{Cassano}, {Brunetti}, {Setti}, {Govoni}  \&
  {Dolag}}{{Cassano} et~al.}{2007}]{2007MNRAS.378.1565C}
{Cassano} R.,  {Brunetti} G.,  {Setti} G.,  {Govoni} F.,   {Dolag} K.,  2007,
  \mn@doi [\mnras] {10.1111/j.1365-2966.2007.11901.x}, \href
  {https://ui.adsabs.harvard.edu/abs/2007MNRAS.378.1565C} {378, 1565}

\bibitem[\protect\citeauthoryear{{Cavagnolo}, {Donahue}, {Voit}  \&
  {Sun}}{{Cavagnolo} et~al.}{2009}]{2009ApJS..182...12C}
{Cavagnolo} K.~W.,  {Donahue} M.,  {Voit} G.~M.,   {Sun} M.,  2009, \mn@doi
  [\apjs] {10.1088/0067-0049/182/1/12}, \href
  {https://ui.adsabs.harvard.edu/abs/2009ApJS..182...12C} {182, 12}

\bibitem[\protect\citeauthoryear{{Condon}, {Cotton}, {Greisen}, {Yin},
  {Perley}, {Taylor}  \& {Broderick}}{{Condon}
  et~al.}{1998}]{1998AJ....115.1693C}
{Condon} J.~J.,  {Cotton} W.~D.,  {Greisen} E.~W.,  {Yin} Q.~F.,  {Perley}
  R.~A.,  {Taylor} G.~B.,   {Broderick} J.~J.,  1998, \mn@doi [\aj]
  {10.1086/300337}, \href
  {https://ui.adsabs.harvard.edu/abs/1998AJ....115.1693C} {115, 1693}

\bibitem[\protect\citeauthoryear{{Conway}, {Cornwell}  \& {Wilkinson}}{{Conway}
  et~al.}{1990}]{1990MNRAS.246..490C}
{Conway} J.~E.,  {Cornwell} T.~J.,   {Wilkinson} P.~N.,  1990, \mnras, \href
  {https://ui.adsabs.harvard.edu/abs/1990MNRAS.246..490C} {246, 490}

\bibitem[\protect\citeauthoryear{{Ebeling}, {Edge}, {Mantz}, {Barrett},
  {Henry}, {Ma}  \& {van Speybroeck}}{{Ebeling}
  et~al.}{2010}]{2010MNRAS.407...83E}
{Ebeling} H.,  {Edge} A.~C.,  {Mantz} A.,  {Barrett} E.,  {Henry} J.~P.,  {Ma}
  C.~J.,   {van Speybroeck} L.,  2010, \mn@doi [\mnras]
  {10.1111/j.1365-2966.2010.16920.x}, \href
  {https://ui.adsabs.harvard.edu/abs/2010MNRAS.407...83E} {407, 83}

\bibitem[\protect\citeauthoryear{{Feretti}, {Giovannini}, {Govoni}  \&
  {Murgia}}{{Feretti} et~al.}{2012}]{2012A&ARv..20...54F}
{Feretti} L.,  {Giovannini} G.,  {Govoni} F.,   {Murgia} M.,  2012, \mn@doi
  [\aapr] {10.1007/s00159-012-0054-z}, \href
  {https://ui.adsabs.harvard.edu/abs/2012A&ARv..20...54F} {20, 54}

\bibitem[\protect\citeauthoryear{{Giacintucci}, {Markevitch}, {Venturi},
  {Clarke}, {Cassano}  \& {Mazzotta}}{{Giacintucci}
  et~al.}{2014a}]{2014ApJ...781....9G}
{Giacintucci} S.,  {Markevitch} M.,  {Venturi} T.,  {Clarke} T.~E.,  {Cassano}
  R.,   {Mazzotta} P.,  2014a, \mn@doi [\apj] {10.1088/0004-637X/781/1/9},
  \href {https://ui.adsabs.harvard.edu/abs/2014ApJ...781....9G} {781, 9}

\bibitem[\protect\citeauthoryear{{Giacintucci}, {Markevitch}, {Brunetti},
  {ZuHone}, {Venturi}, {Mazzotta}  \& {Bourdin}}{{Giacintucci}
  et~al.}{2014b}]{2014ApJ...795...73G}
{Giacintucci} S.,  {Markevitch} M.,  {Brunetti} G.,  {ZuHone} J.~A.,  {Venturi}
  T.,  {Mazzotta} P.,   {Bourdin} H.,  2014b, \mn@doi [\apj]
  {10.1088/0004-637X/795/1/73}, \href
  {https://ui.adsabs.harvard.edu/abs/2014ApJ...795...73G} {795, 73}

\bibitem[\protect\citeauthoryear{{Giacintucci}, {Markevitch}, {Cassano},
  {Venturi}, {Clarke}  \& {Brunetti}}{{Giacintucci}
  et~al.}{2017}]{2017ApJ...841...71G}
{Giacintucci} S.,  {Markevitch} M.,  {Cassano} R.,  {Venturi} T.,  {Clarke}
  T.~E.,   {Brunetti} G.,  2017, \mn@doi [\apj] {10.3847/1538-4357/aa7069},
  \href {https://ui.adsabs.harvard.edu/abs/2017ApJ...841...71G} {841, 71}

\bibitem[\protect\citeauthoryear{{Giacintucci}, {Markevitch}, {Cassano},
  {Venturi}, {Clarke}, {Kale}  \& {Cuciti}}{{Giacintucci}
  et~al.}{2019}]{2019ApJ...880...70G}
{Giacintucci} S.,  {Markevitch} M.,  {Cassano} R.,  {Venturi} T.,  {Clarke}
  T.~E.,  {Kale} R.,   {Cuciti} V.,  2019, \mn@doi [\apj]
  {10.3847/1538-4357/ab29f1}, \href
  {https://ui.adsabs.harvard.edu/abs/2019ApJ...880...70G} {880, 70}

\bibitem[\protect\citeauthoryear{{Giovannini} et~al.,}{{Giovannini}
  et~al.}{2020}]{2020A&A...640A.108G}
{Giovannini} G.,  et~al., 2020, \mn@doi [\aap] {10.1051/0004-6361/202038263},
  \href {https://ui.adsabs.harvard.edu/abs/2020A&A...640A.108G} {640, A108}

\bibitem[\protect\citeauthoryear{{Gitti}, {Brighenti}  \& {McNamara}}{{Gitti}
  et~al.}{2012}]{2012AdAst2012E...6G}
{Gitti} M.,  {Brighenti} F.,   {McNamara} B.~R.,  2012, \mn@doi [Advances in
  Astronomy] {10.1155/2012/950641}, \href
  {https://ui.adsabs.harvard.edu/abs/2012AdAst2012E...6G} {2012, 950641}

\bibitem[\protect\citeauthoryear{{Gitti}, {Brunetti}, {Cassano}  \&
  {Ettori}}{{Gitti} et~al.}{2018}]{2018A&A...617A..11G}
{Gitti} M.,  {Brunetti} G.,  {Cassano} R.,   {Ettori} S.,  2018, \mn@doi [\aap]
  {10.1051/0004-6361/201832749}, \href
  {https://ui.adsabs.harvard.edu/abs/2018A&A...617A..11G} {617, A11}

\bibitem[\protect\citeauthoryear{{Gondhalekar}, {Saha}, {Safonova}  \&
  {Mathur}}{{Gondhalekar} et~al.}{2022}]{2022arXiv220710973G}
{Gondhalekar} Y.,  {Saha} S.,  {Safonova} M.,   {Mathur} A.,  2022, arXiv
  e-prints, \href {https://ui.adsabs.harvard.edu/abs/2022arXiv220710973G} {p.
  arXiv:2207.10973}

\bibitem[\protect\citeauthoryear{{Govoni}, {Murgia}, {Markevitch}, {Feretti},
  {Giovannini}, {Taylor}  \& {Carretti}}{{Govoni}
  et~al.}{2009}]{2009A&A...499..371G}
{Govoni} F.,  {Murgia} M.,  {Markevitch} M.,  {Feretti} L.,  {Giovannini} G.,
  {Taylor} G.~B.,   {Carretti} E.,  2009, \mn@doi [\aap]
  {10.1051/0004-6361/200811180}, \href
  {https://ui.adsabs.harvard.edu/abs/2009A&A...499..371G} {499, 371}

\bibitem[\protect\citeauthoryear{Grobler, Nunhokee, Smirnov, van Zyl  \& de
  Bruyn}{Grobler et~al.}{2014}]{10.1093/mnras/stu268}
Grobler T.~L.,  Nunhokee C.~D.,  Smirnov O.~M.,  van Zyl A.~J.,   de Bruyn
  A.~G.,  2014, \mn@doi [Monthly Notices of the Royal Astronomical Society]
  {10.1093/mnras/stu268}, 439, 4030

\bibitem[\protect\citeauthoryear{Harwood, Hardcastle, Croston  \&
  Goodger}{Harwood et~al.}{2013}]{10.1093/mnras/stt1526}
Harwood J.~J.,  Hardcastle M.~J.,  Croston J.~H.,   Goodger J.~L.,  2013,
  \mn@doi [Monthly Notices of the Royal Astronomical Society]
  {10.1093/mnras/stt1526}, 435, 3353

\bibitem[\protect\citeauthoryear{Harwood, Hardcastle  \& Croston}{Harwood
  et~al.}{2015}]{10.1093/mnras/stv2194}
Harwood J.~J.,  Hardcastle M.~J.,   Croston J.~H.,  2015, \mn@doi [Monthly
  Notices of the Royal Astronomical Society] {10.1093/mnras/stv2194}, 454, 3403

\bibitem[\protect\citeauthoryear{{Hugo}, {Perkins}, {Merry}, {Mauch}  \&
  {Smirnov}}{{Hugo} et~al.}{2022}]{2022arXiv220609179H}
{Hugo} B.~V.,  {Perkins} S.,  {Merry} B.,  {Mauch} T.,   {Smirnov} O.~M.,
  2022, arXiv e-prints, \href
  {https://ui.adsabs.harvard.edu/abs/2022arXiv220609179H} {p. arXiv:2206.09179}

\bibitem[\protect\citeauthoryear{{Ignesti} et~al.,}{{Ignesti}
  et~al.}{2022}]{2022A&A...659A..20I}
{Ignesti} A.,  et~al., 2022, \mn@doi [\aap] {10.1051/0004-6361/202142549},
  \href {https://ui.adsabs.harvard.edu/abs/2022A&A...659A..20I} {659, A20}

\bibitem[\protect\citeauthoryear{{Iheanetu}, {Girard}, {Smirnov}, {Asad}, {de
  Villiers}, {Thorat}, {Makhathini}  \& {Perley}}{{Iheanetu}
  et~al.}{2019}]{2019MNRAS.485.4107I}
{Iheanetu} K.,  {Girard} J.~N.,  {Smirnov} O.,  {Asad} K.~M.~B.,  {de Villiers}
  M.,  {Thorat} K.,  {Makhathini} S.,   {Perley} R.~A.,  2019, \mn@doi [\mnras]
  {10.1093/mnras/stz702}, \href
  {https://ui.adsabs.harvard.edu/abs/2019MNRAS.485.4107I} {485, 4107}

\bibitem[\protect\citeauthoryear{{Jacob} \& {Pfrommer}}{{Jacob} \&
  {Pfrommer}}{2017a}]{2017MNRAS.467.1449J}
{Jacob} S.,  {Pfrommer} C.,  2017a, \mn@doi [\mnras] {10.1093/mnras/stx131},
  \href {https://ui.adsabs.harvard.edu/abs/2017MNRAS.467.1449J} {467, 1449}

\bibitem[\protect\citeauthoryear{{Jacob} \& {Pfrommer}}{{Jacob} \&
  {Pfrommer}}{2017b}]{2017MNRAS.467.1478J}
{Jacob} S.,  {Pfrommer} C.,  2017b, \mn@doi [\mnras] {10.1093/mnras/stx132},
  \href {https://ui.adsabs.harvard.edu/abs/2017MNRAS.467.1478J} {467, 1478}

\bibitem[\protect\citeauthoryear{{Jonas} \& {MeerKAT Team}}{{Jonas} \& {MeerKAT
  Team}}{2016}]{2016mks..confE...1J}
{Jonas} J.,  {MeerKAT Team} 2016, in MeerKAT Science: On the Pathway to the
  SKA. p.~1

\bibitem[\protect\citeauthoryear{{J{\'o}zsa} et~al.,}{{J{\'o}zsa}
  et~al.}{2020}]{2020ascl.soft06014J}
{J{\'o}zsa} G. I.~G.,  et~al., 2020, {CARACal: Containerized Automated Radio
  Astronomy Calibration pipeline} (\mn@eprint {ascl} {2006.014})

\bibitem[\protect\citeauthoryear{{Kenyon}, {Smirnov}, {Grobler}  \&
  {Perkins}}{{Kenyon} et~al.}{2018}]{2018MNRAS.478.2399K}
{Kenyon} J.~S.,  {Smirnov} O.~M.,  {Grobler} T.~L.,   {Perkins} S.~J.,  2018,
  \mn@doi [\mnras] {10.1093/mnras/sty1221}, \href
  {https://ui.adsabs.harvard.edu/abs/2018MNRAS.478.2399K} {478, 2399}

\bibitem[\protect\citeauthoryear{{Knowles} et~al.,}{{Knowles}
  et~al.}{2022}]{2022A&A...657A..56K}
{Knowles} K.,  et~al., 2022, \mn@doi [\aap] {10.1051/0004-6361/202141488},
  \href {https://ui.adsabs.harvard.edu/abs/2022A&A...657A..56K} {657, A56}

\bibitem[\protect\citeauthoryear{{Kokotanekov}, {Wise}, {de Vries}  \&
  {Intema}}{{Kokotanekov} et~al.}{2018}]{2018A&A...618A.152K}
{Kokotanekov} G.,  {Wise} M.~W.,  {de Vries} M.,   {Intema} H.~T.,  2018,
  \mn@doi [\aap] {10.1051/0004-6361/201833222}, \href
  {https://ui.adsabs.harvard.edu/abs/2018A&A...618A.152K} {618, A152}

\bibitem[\protect\citeauthoryear{Makhathini}{Makhathini}{2018}]{makhathini2018}
Makhathini S.,  2018, PhD thesis, Rhodes University, Drosty Rd, Grahamstown,
  6139, Eastern Cape, South Africa

\bibitem[\protect\citeauthoryear{{McMullin}, {Waters}, {Schiebel}, {Young}  \&
  {Golap}}{{McMullin} et~al.}{2007}]{2007ASPC..376..127M}
{McMullin} J.~P.,  {Waters} B.,  {Schiebel} D.,  {Young} W.,   {Golap} K.,
  2007, in {Shaw} R.~A.,  {Hill} F.,   {Bell} D.~J.,  eds,  Astronomical
  Society of the Pacific Conference Series Vol. 376, Astronomical Data Analysis
  Software and Systems XVI. p.~127

\bibitem[\protect\citeauthoryear{{Mohan} \& {Rafferty}}{{Mohan} \&
  {Rafferty}}{2015}]{2015ascl.soft02007M}
{Mohan} N.,  {Rafferty} D.,  2015, {PyBDSF: Python Blob Detection and Source
  Finder}, Astrophysics Source Code Library, record ascl:1502.007 (\mn@eprint
  {ascl} {1502.007})

\bibitem[\protect\citeauthoryear{{Noordam} \& {Smirnov}}{{Noordam} \&
  {Smirnov}}{2010}]{2010A&A...524A..61N}
{Noordam} J.~E.,  {Smirnov} O.~M.,  2010, \mn@doi [\aap]
  {10.1051/0004-6361/201015013}, \href
  {https://ui.adsabs.harvard.edu/abs/2010A&A...524A..61N} {524, A61}

\bibitem[\protect\citeauthoryear{{Offringa} \& {Smirnov}}{{Offringa} \&
  {Smirnov}}{2017}]{2017MNRAS.471..301O}
{Offringa} A.~R.,  {Smirnov} O.,  2017, \mn@doi [\mnras]
  {10.1093/mnras/stx1547}, \href
  {https://ui.adsabs.harvard.edu/abs/2017MNRAS.471..301O} {471, 301}

\bibitem[\protect\citeauthoryear{{Offringa} et~al.,}{{Offringa}
  et~al.}{2014}]{2014MNRAS.444..606O}
{Offringa} A.~R.,  et~al., 2014, \mn@doi [\mnras] {10.1093/mnras/stu1368},
  \href {https://ui.adsabs.harvard.edu/abs/2014MNRAS.444..606O} {444, 606}

\bibitem[\protect\citeauthoryear{{Pandey-Pommier}, {Richard}, {Combes}, {Edge},
  {Guiderdoni}, {Narasimha}, {Bagchi}  \& {Jacob}}{{Pandey-Pommier}
  et~al.}{2016}]{2016sf2a.conf..367P}
{Pandey-Pommier} M.,  {Richard} J.,  {Combes} F.,  {Edge} A.,  {Guiderdoni} B.,
   {Narasimha} D.,  {Bagchi} J.,   {Jacob} J.,  2016, in {Reyl{\'e}} C.,
  {Richard} J.,  {Cambr{\'e}sy} L.,  {Deleuil} M.,  {P{\'e}contal} E.,
  {Tresse} L.,   {Vauglin} I.,  eds, SF2A-2016: Proceedings of the Annual
  meeting of the French Society of Astronomy and Astrophysics. pp 367--372
  (\mn@eprint {arXiv} {1612.00225})

\bibitem[\protect\citeauthoryear{{Perley} \& {Butler}}{{Perley} \&
  {Butler}}{2017}]{2017ApJS..230....7P}
{Perley} R.~A.,  {Butler} B.~J.,  2017, \mn@doi [\apjs]
  {10.3847/1538-4365/aa6df9}, \href
  {https://ui.adsabs.harvard.edu/abs/2017ApJS..230....7P} {230, 7}

\bibitem[\protect\citeauthoryear{{Perrott}, {SM}, {Edge}, {Grainge}, {Green}
  \& {Saunders}}{{Perrott} et~al.}{2023}]{2023MNRAS.520L...1P}
{Perrott} Y.~C.,  {SM} G.,  {Edge} A.~C.,  {Grainge} K. J.~B.,  {Green} D.~A.,
   {Saunders} R. D.~E.,  2023, \mn@doi [\mnras] {10.1093/mnrasl/slac160}, \href
  {https://ui.adsabs.harvard.edu/abs/2023MNRAS.520L...1P} {520, L1}

\bibitem[\protect\citeauthoryear{{Piffaretti}, {Arnaud}, {Pratt},
  {Pointecouteau}  \& {Melin}}{{Piffaretti} et~al.}{2011}]{2011A&A...534A.109P}
{Piffaretti} R.,  {Arnaud} M.,  {Pratt} G.~W.,  {Pointecouteau} E.,   {Melin}
  J.~B.,  2011, \mn@doi [\aap] {10.1051/0004-6361/201015377}, \href
  {https://ui.adsabs.harvard.edu/abs/2011A&A...534A.109P} {534, A109}

\bibitem[\protect\citeauthoryear{{Planck Collaboration} et~al.,}{{Planck
  Collaboration} et~al.}{2014}]{2014A&A...571A..29P}
{Planck Collaboration} et~al., 2014, \mn@doi [\aap]
  {10.1051/0004-6361/201321523}, \href
  {https://ui.adsabs.harvard.edu/abs/2014A&A...571A..29P} {571, A29}

\bibitem[\protect\citeauthoryear{{Richard-Laferri{\`e}re}
  et~al.,}{{Richard-Laferri{\`e}re} et~al.}{2020}]{2020MNRAS.499.2934R}
{Richard-Laferri{\`e}re} A.,  et~al., 2020, \mn@doi [\mnras]
  {10.1093/mnras/staa2877}, \href
  {https://ui.adsabs.harvard.edu/abs/2020MNRAS.499.2934R} {499, 2934}

\bibitem[\protect\citeauthoryear{{Riseley} et~al.,}{{Riseley}
  et~al.}{2022}]{2022MNRAS.512.4210R}
{Riseley} C.~J.,  et~al., 2022, \mn@doi [\mnras] {10.1093/mnras/stac672}, \href
  {https://ui.adsabs.harvard.edu/abs/2022MNRAS.512.4210R} {512, 4210}

\bibitem[\protect\citeauthoryear{{Sault} \& {Conway}}{{Sault} \&
  {Conway}}{1999}]{1999ASPC..180..419S}
{Sault} R.~J.,  {Conway} J.~E.,  1999, in {Taylor} G.~B.,  {Carilli} C.~L.,
  {Perley} R.~A.,  eds,  Astronomical Society of the Pacific Conference Series
  Vol. 180, Synthesis Imaging in Radio Astronomy II. p.~419

\bibitem[\protect\citeauthoryear{{Savini} et~al.,}{{Savini}
  et~al.}{2019}]{2019A&A...622A..24S}
{Savini} F.,  et~al., 2019, \mn@doi [\aap] {10.1051/0004-6361/201833882}, \href
  {https://ui.adsabs.harvard.edu/abs/2019A&A...622A..24S} {622, A24}

\bibitem[\protect\citeauthoryear{{Smirnov}}{{Smirnov}}{2011a}]{2011A&A...527A.106S}
{Smirnov} O.~M.,  2011a, \mn@doi [\aap] {10.1051/0004-6361/201016082}, \href
  {https://ui.adsabs.harvard.edu/abs/2011A&A...527A.106S} {527, A106}

\bibitem[\protect\citeauthoryear{{Smirnov}}{{Smirnov}}{2011b}]{2011A&A...527A.107S}
{Smirnov} O.~M.,  2011b, \mn@doi [\aap] {10.1051/0004-6361/201116434}, \href
  {https://ui.adsabs.harvard.edu/abs/2011A&A...527A.107S} {527, A107}

\bibitem[\protect\citeauthoryear{{Smirnov}}{{Smirnov}}{2011c}]{2011A&A...527A.108S}
{Smirnov} O.~M.,  2011c, \mn@doi [\aap] {10.1051/0004-6361/201116435}, \href
  {https://ui.adsabs.harvard.edu/abs/2011A&A...527A.108S} {527, A108}

\bibitem[\protect\citeauthoryear{{Smirnov}}{{Smirnov}}{2011d}]{2011A&A...531A.159S}
{Smirnov} O.~M.,  2011d, \mn@doi [\aap] {10.1051/0004-6361/201116764}, \href
  {https://ui.adsabs.harvard.edu/abs/2011A&A...531A.159S} {531, A159}

\bibitem[\protect\citeauthoryear{{Smirnov} \& {Tasse}}{{Smirnov} \&
  {Tasse}}{2015}]{2015MNRAS.449.2668S}
{Smirnov} O.~M.,  {Tasse} C.,  2015, \mn@doi [\mnras] {10.1093/mnras/stv418},
  \href {https://ui.adsabs.harvard.edu/abs/2015MNRAS.449.2668S} {449, 2668}

\bibitem[\protect\citeauthoryear{Sob, Bester, Smirnov, Kenyon  \& Grobler}{Sob
  et~al.}{2019}]{10.1093/mnras/stz3037}
Sob U.~M.,  Bester H.~L.,  Smirnov O.~M.,  Kenyon J.~S.,   Grobler T.~L.,
  2019, \mn@doi [Monthly Notices of the Royal Astronomical Society]
  {10.1093/mnras/stz3037}, 491, 1026

\bibitem[\protect\citeauthoryear{{Tasse}}{{Tasse}}{2014a}]{2014arXiv1410.8706T}
{Tasse} C.,  2014a, arXiv e-prints, \href
  {https://ui.adsabs.harvard.edu/abs/2014arXiv1410.8706T} {p. arXiv:1410.8706}

\bibitem[\protect\citeauthoryear{{Tasse}}{{Tasse}}{2014b}]{2014A&A...566A.127T}
{Tasse} C.,  2014b, \mn@doi [\aap] {10.1051/0004-6361/201423503}, \href
  {https://ui.adsabs.harvard.edu/abs/2014A&A...566A.127T} {566, A127}

\bibitem[\protect\citeauthoryear{{Tasse} et~al.,}{{Tasse}
  et~al.}{2018}]{2018A&A...611A..87T}
{Tasse} C.,  et~al., 2018, \mn@doi [\aap] {10.1051/0004-6361/201731474}, \href
  {https://ui.adsabs.harvard.edu/abs/2018A&A...611A..87T} {611, A87}

\bibitem[\protect\citeauthoryear{{Venturi}, {Giacintucci}, {Brunetti},
  {Cassano}, {Bardelli}, {Dallacasa}  \& {Setti}}{{Venturi}
  et~al.}{2007}]{2007A&A...463..937V}
{Venturi} T.,  {Giacintucci} S.,  {Brunetti} G.,  {Cassano} R.,  {Bardelli} S.,
   {Dallacasa} D.,   {Setti} G.,  2007, \mn@doi [\aap]
  {10.1051/0004-6361:20065961}, \href
  {https://ui.adsabs.harvard.edu/abs/2007A&A...463..937V} {463, 937}

\bibitem[\protect\citeauthoryear{Wijnholds, Grobler  \& Smirnov}{Wijnholds
  et~al.}{2016}]{10.1093/mnras/stw118}
Wijnholds S.~J.,  Grobler T.~L.,   Smirnov O.~M.,  2016, \mn@doi [Monthly
  Notices of the Royal Astronomical Society] {10.1093/mnras/stw118}, 457, 2331

\bibitem[\protect\citeauthoryear{{Yu} et~al.,}{{Yu}
  et~al.}{2018}]{2018ApJ...853..100Y}
{Yu} H.,  et~al., 2018, \mn@doi [\apj] {10.3847/1538-4357/aaa421}, \href
  {https://ui.adsabs.harvard.edu/abs/2018ApJ...853..100Y} {853, 100}

\bibitem[\protect\citeauthoryear{{ZuHone}, {Markevitch}, {Brunetti}  \&
  {Giacintucci}}{{ZuHone} et~al.}{2013}]{2013ApJ...762...78Z}
{ZuHone} J.~A.,  {Markevitch} M.,  {Brunetti} G.,   {Giacintucci} S.,  2013,
  \mn@doi [\apj] {10.1088/0004-637X/762/2/78}, \href
  {https://ui.adsabs.harvard.edu/abs/2013ApJ...762...78Z} {762, 78}

\bibitem[\protect\citeauthoryear{{ZuHone}, {Brunetti}, {Giacintucci}  \&
  {Markevitch}}{{ZuHone} et~al.}{2015}]{2015ApJ...801..146Z}
{ZuHone} J.~A.,  {Brunetti} G.,  {Giacintucci} S.,   {Markevitch} M.,  2015,
  \mn@doi [\apj] {10.1088/0004-637X/801/2/146}, \href
  {https://ui.adsabs.harvard.edu/abs/2015ApJ...801..146Z} {801, 146}

\bibitem[\protect\citeauthoryear{{de Villiers} \& {Cotton}}{{de Villiers} \&
  {Cotton}}{2022}]{2022AJ....163..135D}
{de Villiers} M.~S.,  {Cotton} W.~D.,  2022, \mn@doi [\aj]
  {10.3847/1538-3881/ac460a}, \href
  {https://ui.adsabs.harvard.edu/abs/2022AJ....163..135D} {163, 135}

\bibitem[\protect\citeauthoryear{{van Weeren}, {de Gasperin}, {Akamatsu},
  {Br{\"u}ggen}, {Feretti}, {Kang}, {Stroe}  \& {Zandanel}}{{van Weeren}
  et~al.}{2019}]{2019SSRv..215...16V}
{van Weeren} R.~J.,  {de Gasperin} F.,  {Akamatsu} H.,  {Br{\"u}ggen} M.,
  {Feretti} L.,  {Kang} H.,  {Stroe} A.,   {Zandanel} F.,  2019, \mn@doi [\ssr]
  {10.1007/s11214-019-0584-z}, \href
  {https://ui.adsabs.harvard.edu/abs/2019SSRv..215...16V} {215, 16}

\makeatother
\end{thebibliography}

	%%%%%%%%%%%%%%%%%%%%%%%%%%%%%%%%%%%%%%%%%%%%%%%%%%
	
	%%%%%%%%%%%%%%%%% APPENDICES %%%%%%%%%%%%%%%%%%%%%
	
	\appendix
	
	\section{Abbreviations} \label{appendix: abbreviations}
	
	List of abbreviations used in this work.
	
	\begin{table*}
		\centering
		\caption{List of specialised abbreviations. Columns: contraction, full phrase.}
		\begin{tabular}{ll|ll}
			\hline
			Contraction                & Phrase                    & Contraction & Phrase\\
			\hline
			1GC & First generation calibration   & lsm & local sky model \\
			2GC & Second generation calibration & kMS & killMS   \\
			3GC & Third generation calibration & DDF & DDFacet \\
			DIE & Direction independent effect & SSD & Sub-Space Deconvolution \\
			DDE & Direction dependent effect & HR & High resolution \\ 
			DR & Dynamic range & LR & Low resolution \\
			KATDAL & Karoo Array Telescope Data Access Library &  SRC-SUB & Source subtracted \\
			CARACal & Containerized Automated Radio Astronomy Calibration & LLS & Largest linear size \\
			\hline
		\end{tabular}
		\label{tab:Table A.1}
	\end{table*}
	
	\section{Software} \label{appendix: software}
	List of software used in this work.
	
	\begin{table*}
		\centering
		\caption{Software links. Columns: software, web link.}
		\begin{tabular}{ll|ll}
			\hline
			Software                & Link                     & Software & Link\\
			\hline
			KATDAL       & \url{https://github.com/ska-sa/katdal} &     DDFacet        & \url{https://github.com/saopicc/DDFacet} \\
			CARACal          & \url{https://github.com/caracal-pipeline/caracal} &      eidos       & \url{https://github.com/ratt-ru/eidos} \\
			Tricolour        & \url{https://github.com/ska-sa/tricolour} &     breizorro        & \url{https://github.com/ratt-ru/breizorro} \\
			WSClean         &  \url{https://gitlab.com/aroffringa/wsclean} &     Stimela        & \url{https://github.com/ratt-ru/Stimela}\\
			CubiCal         & \url{https://github.com/ratt-ru/CubiCal} 	&     brats         & \url{http://www.askanastronomer.co.uk/brats}\\
			killMS         & \url{https://github.com/saopicc/killMS} & QuartiCal & \url{https://github.com/ratt-ru/QuartiCal} \\
			\hline
		\end{tabular}
		\label{tab:Table B.1}
	\end{table*}
	
	\section{ACO 1795 images} \label{appendix: 1795}
	
\begin{figure*}
		\centering
		\includegraphics[width=\textwidth]{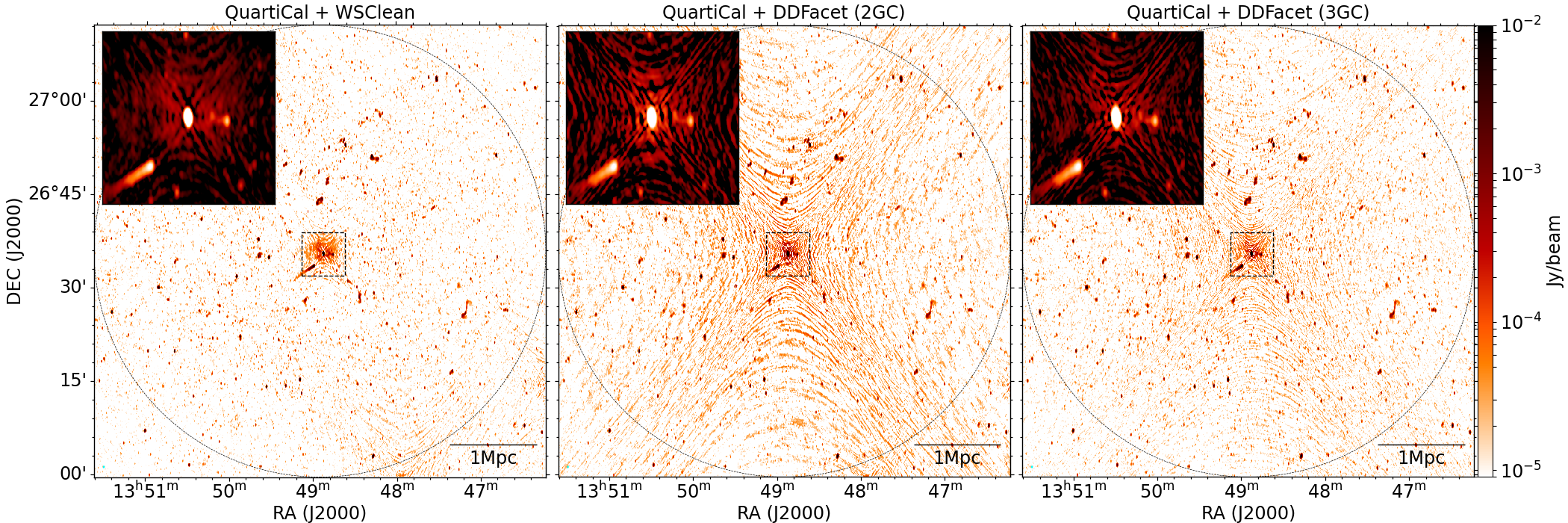}
		\caption{ACO 1795 images. \textit{Left:} best 2GC image we could produce, where QuartiCal self-calibrated visibilities were imaged with WSClean for three rounds of chained delay and complex-gain solutions. \textit{Middle:} visibilities from the \textit{Left} imaged with DDFacet as per described in Section \ref{section:3GC} until Step 3. \textit{Right:} visibilities from the \textit{Middle} after 3GC kMS solutions have been applied similar to the other fields. The black dashed circle shows the MeerKAT primary beam and the black dashed square indicates the central 500 kpc box around the BCG as shown by the insets. All images are on the same colorscale and convolved to a (15$\arcsec$, 7$\arcsec$, 0$^{\circ}$). The global rms noise from left to right is: 6.7, 11.2, 8.7 $\mu$Jy/beam, respectively. The local rms noise in each inset from left to right is: 34.9, 91.0, 58.5 $\mu$Jy/beam, respectively.}
		\label{fig:1795}
	\end{figure*}

	Figure \ref{fig:1795} shows our highest fidelity images of ACO 1795, each of which highlights the BCG and its strong artefacts obscuring the cluster core. We suspect these artefacts are not direction-dependent in nature and are in fact due to polarisation leakage effects, compounded by a very asymmetric PSF, as described in Section \ref{section:1795}. The highest fidelity image (with the least artefacts and lowest noise) was produced with QuartiCal and WSClean. The 3GC solutions reduced the artefacts in the corresponding DDF 2GC image similar to that seen in ACO 1413, but could not improve upon the WSClean image. A detailed comparison between the two imagers is outside the scope of this work.
	
	\section{2GC and 3GC Field of Views} \label{appendix: 2vs3gc}

\begin{figure*}
		%\vspace{1cm}
		\centering
		\includegraphics[width=\textwidth]{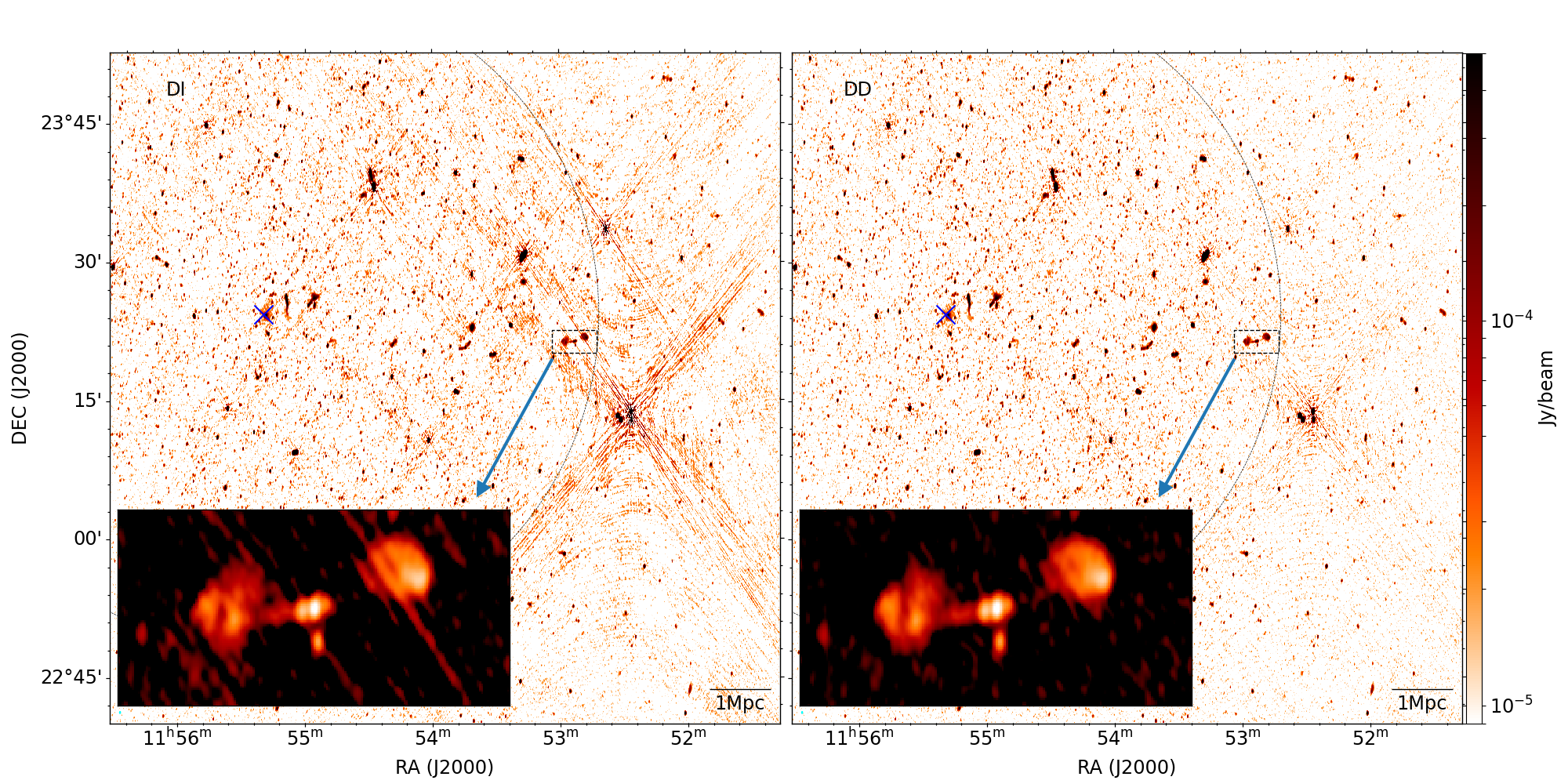}
		\caption{ACO 1413 2GC vs 3GC artefacts. Left: 2GC image, beam size (12.3$\arcsec$, 5.9$\arcsec$, -4.6$^{\circ}$), global rms noise 8.3 $\mu$Jy/beam. Right: 3GC image, beam (12.2$\arcsec$, 5.9$\arcsec$, -4.5$^{\circ}$), global rms noise 7.4 $\mu$Jy/beam. A blue cross marks the phase centre of the images. The thin dashed black arc indicates the primary beam of the telescope. The bold dashed black rectangle marks a FR-II galaxy that is contaminated by an artefact with an arrow pointing to a zoomed inset of this source displayed in the inverse colorbar. The local rms of this source decreased by just under a factor of two in the 2GC to 3GC images.}
		\label{fig:A1413_artifacts}
	\end{figure*}

\begin{figure*}
		%\vspace{1cm}
		\centering
		\includegraphics[width=\textwidth]{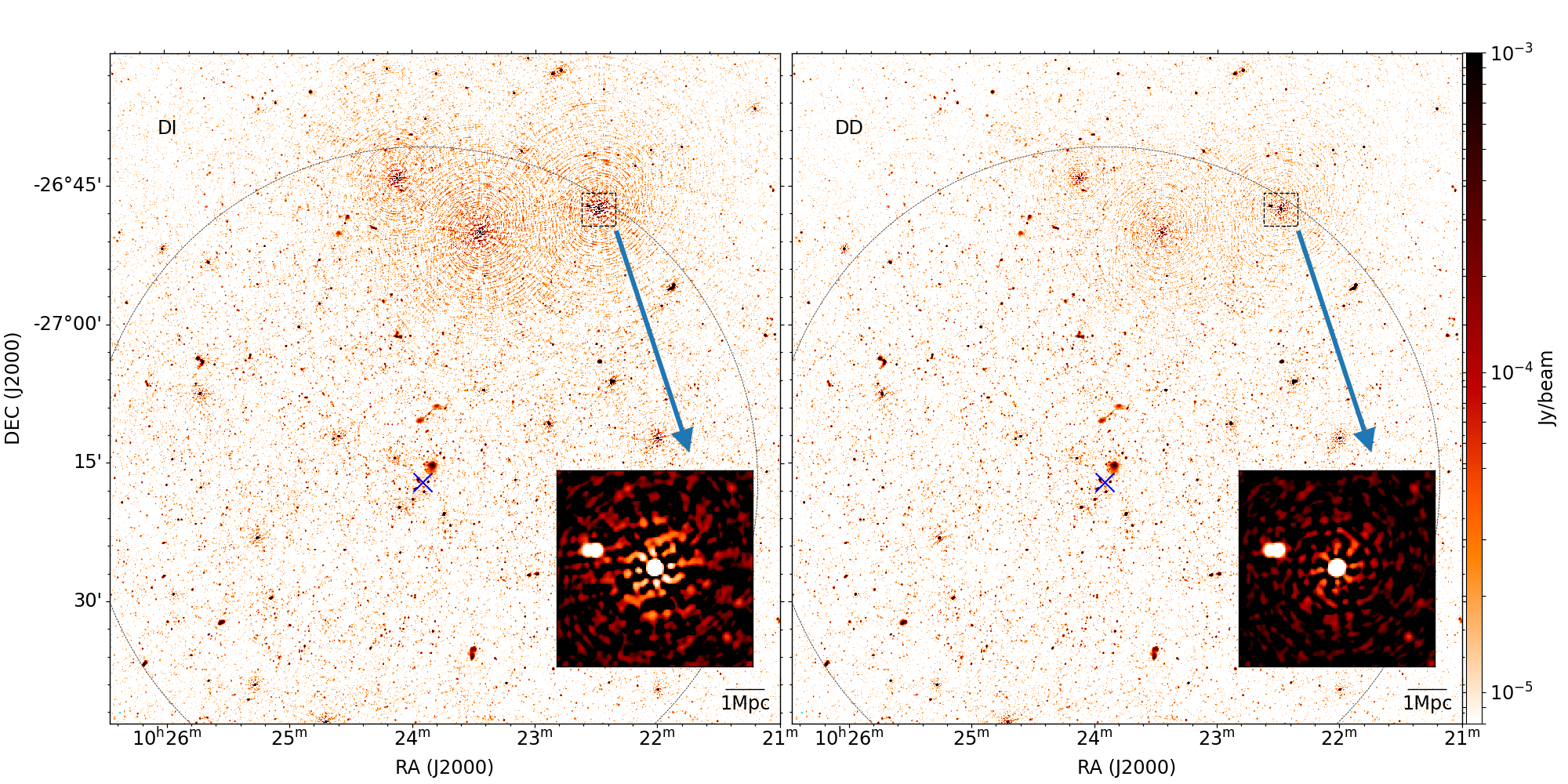}
		\caption{ACO 3444 2GC vs 3GC artefacts. Left: 2GC image, beam size (7.4$\arcsec$, 6.0$\arcsec$, -10.3$^{\circ}$), global rms noise 6.4 $\mu$Jy/beam. Right: 3GC image, beam (6.8$\arcsec$, 5.9$\arcsec$, -16.0$^{\circ}$), global rms noise 5.9 $\mu$Jy/beam. A blue cross marks the phase centre of the images. The thin dashed black arc indicates the primary beam of the telescope. The bold dashed black rectangle marks a bright point source with an arrow pointing to a zoomed inset of this source displayed in the inverse colorbar. The local rms decreased by just over a factor of two.}
		\label{fig:A3444_artifacts}
	\end{figure*}

\begin{figure*}
		\centering
		\includegraphics[width=\textwidth]{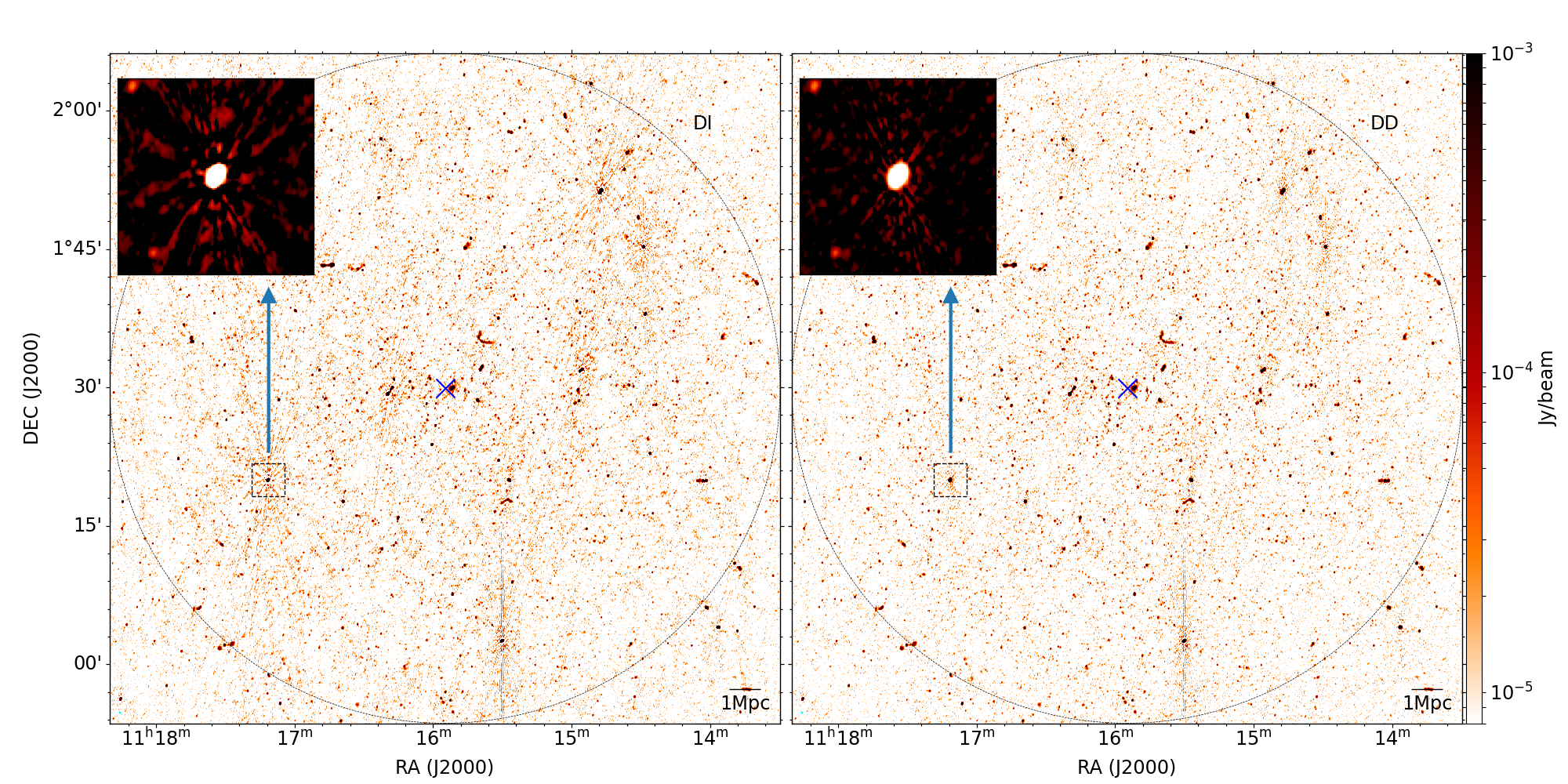}
		\caption{MACS J1115.8+0129 2GC vs 3GC artefacts. Left: 2GC image, beam size (8.6$\arcsec$, 6.6$\arcsec$, -16.7$^{\circ}$), global apparent rms noise 8.2 $\mu$Jy/beam. Right: 3GC image, beam (8.5$\arcsec$, 6.5$\arcsec$, -16.2$^{\circ}$), global rms noise 7.6 $\mu$Jy/beam. A blue cross marks the phase centre of the images. The thin dashed black arc indicates the primary beam of the telescope. The bold dashed black rectangle marks a bright point source with an arrow pointing to a zoomed inset of this source displayed in the inverse colorbar. The local rms decreased by just under a factor of two.}
		\label{fig:MACSJ1115.8+0129_artifacts}
	\end{figure*}

\begin{figure*}
		\centering
		\includegraphics[width=\textwidth]{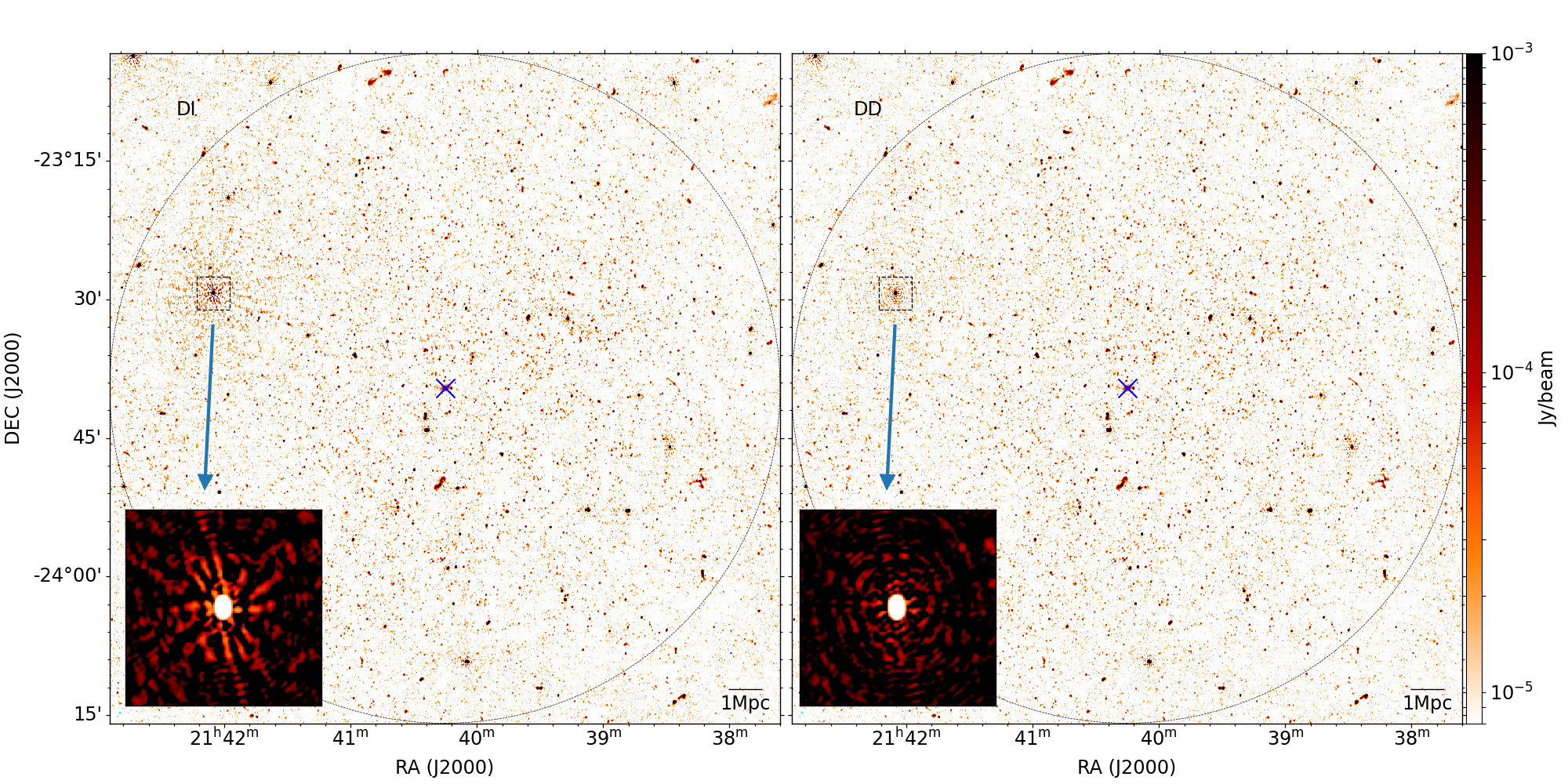}
		\caption{MACS J2140.2-2339 2GC vs 3GC artefacts. Left: 2GC image, beam size (7.9$\arcsec$, 6.3$\arcsec$, -9.7$^{\circ}$), global apparent rms 7.2 $\mu$Jy/beam. Right: 3GC image, beam (7.8$\arcsec$, 6.2$\arcsec$, -8.6$^{\circ}$), global rms noise 7.1 $\mu$Jy/beam. A blue cross marks the phase centre of the images. The thin dashed black arc indicates the primary beam of the telescope. The bold dashed black rectangle marks a bright point source with an arrow pointing to a zoomed inset of this source displayed in the inverse colorbar. The local rms decreased by just under a factor of two.}
		\label{fig:MACSJ2140.2-2339_artifacts}
	\end{figure*}

	Table \ref{tab:Table D.1} gives the 2GC and 3GC global noise and DR values for each cluster except ACO 1795. Figures \ref{fig:A1413_artifacts}-\ref{fig:MACSJ2140.2-2339_artifacts} show comparisons between the 2GC and 3GC field artefacts for each observation except ACO 1795, centred at a position that highlights the artefacts with respect to the phase centre. If no artefacts require focus, the image is centred on the phase centre. The colorscale is the same for each pair of images.
	
	\begin{table*}
		\centering
		\caption{Image properties. Columns: Image corresponding to each cluster, global root mean square, and global dynamic ranges (DRs) calculated with respect to the root mean square and minimum pixel values of the unsubtracted images.}
		\begin{tabular}{ *{7}{c} }
			\hline
			Image    & \multicolumn{2}{c}{\makecell{RMS \\ ($\mu$Jy/beam)}} & \multicolumn{2}{c}{\makecell{DR$^{(1)}$ \\ (max/rms)}}  & \multicolumn{2}{c}{\makecell{DR$^{(2)}$ \\ $\|$max/min$\|$}}       \\
			\hline
			&   2GC  &   3GC  &   2GC  &   3GC  &   2GC  &   3GC   \\
			\hline
			ACO 1413 &  8.3   & 7.4 & 10024 & 11284 & 87 & 354     \\
			SRC-SUB & 8.5 & 7.9 &   &    &    &       \\
			\hline
			ACO 3444  & 6.4  & 5.9 & 26828 & 29017 & 110 & 319   \\ 
			SRC-SUB & 6.8 & 6.1 &   &    &    &       \\  
			\hline
			MACS J1115.8+0129 &   8.2 & 7.6 & 4902 & 5224 & 92 & 265      \\  
			SRC-SUB & 9.1 & 7.9 &   &    &    &       \\  
			\hline  
			MACS J2140.2-2339  & 7.2 & 7.1 & 12208 & 12394 & 198 & 386      \\
			SRC-SUB & 7.3 & 7.1 &   &    &    &       \\  
			\hline
		\end{tabular}
		\label{tab:Table D.1}
	\end{table*}
	
	%%%%%%%%%%%%%%%%%%%%%%%%%%%%%%%%%%%%%%%%%%%%%%%%%%

	% Don't change these lines
	\bsp	% typesetting comment
	\label{lastpage}
\end{document}